\documentclass[11pt]{article}

\usepackage{comment}
\usepackage{booktabs}
\usepackage{threeparttable}

\usepackage{mathrsfs}
\usepackage{fancyhdr}
\usepackage{epsfig,morefloats}
\usepackage[authoryear]{natbib}
\setcitestyle{aysep={,}}
\usepackage{amsmath,enumerate,amsbsy,amsfonts,amssymb,mathabx,amscd,graphicx,algorithm,amsthm,soul,algpseudocode, indentfirst}
\usepackage{adjustbox}
\usepackage{epsfig}
\usepackage{color}
\usepackage{lscape}
\usepackage{rotating}
\usepackage{caption}
\usepackage{subfigure}
\usepackage[compact]{titlesec}
 \usepackage[breaklinks,colorlinks,linkcolor=blue,citecolor=blue,urlcolor=black]{hyperref}
\usepackage{lineno}
\usepackage{cancel}
\usepackage{multirow}
\usepackage{setspace}

\definecolor{mygray}{gray}{0.6}

\newcommand{\Date}[1]{\def\@Date{#1}}
\def\today{\number\day~\ifcase\month\or
 January\or February\or March\or April\or May\or June\or
 July\or August\or September\or October\or November\or December\fi~\number\year}

\def\t0{\theta_0}

\usepackage{graphicx}
\graphicspath{{Pictures/}}

\newcommand{\argmin}{\ensuremath{\operatornamewithlimits{argmin}}}

\def\bsc{\begin{scriptsize}}
\def\esc{\end{scriptsize}}

\newtheorem{theorem}{Theorem}

\newtheorem{lemma}{Lemma}

\theoremstyle{definition}

\newtheorem{remark}{Remark}
\newtheorem{condition}{Condition}

\makeatletter
\newcommand{\figcaption}{\def\@captype{figure}\caption}
\newcommand{\tabcaption}{\def\@captype{table}\caption}
\makeatother

\newcommand{\bSigma}{\boldsymbol \Sigma}

\newcommand{\blind}{1}

\newcommand{\bepsilon}{\boldsymbol \epsilon}

\newcommand{\bPhi}{\boldsymbol \Phi}
\newcommand{\bmu}{\boldsymbol \mu}

\def\p{{ \mathrm{p} }}

\newcommand{\be}{{\mathbf e}}
\newcommand{\bbf}{{\mathbf f}}

\newcommand{\bz}{{\mathbf z}}

\newcommand{\by}{{\mathbf y}}
\newcommand{\bu}{{\mathbf u}}
\newcommand{\bv}{{\mathbf v}}

\newcommand{\bw}{{\mathbf w}}

\newcommand{\bbb}{{\mathbf b}}

\newcommand{\bM}{{\bf M}}
\newcommand{\bD}{{\bf D}}
\newcommand{\bA}{{\bf A}}
\newcommand{\bB}{{\bf B}}
\newcommand{\bC}{{\bf C}}

\newcommand{\bI}{{\bf I}}

\newcommand{\bP}{{\bf P}}

\newcommand{\bX}{{\bf X}}

\newcommand{\eP}{\mathbb{P}}
\newcommand{\E}{\mathbb{E}}

\newcommand{\tF}{\text{F}}

\newcommand{\T}{\mathrm{\scriptscriptstyle T}}

\allowdisplaybreaks

\numberwithin{remark}{section}

\usepackage{geometry}
\usepackage{authblk}

\begin{document}

\def\spacingset#1{\renewcommand{\baselinestretch}%
{#1}\small\normalsize} \spacingset{1}

%\renewcommand{\baselinestretch}{1.0}

%%%%%%%%%%%%%%%%%%%%%%%%%%%%%%%%%%%%%%%%%%%%%%%%%%%%%%%%%%%%%%%%%%%%%%%%%%%%%%

\if1\blind
{
  \title{\bf Large covariance matrix estimation with factor-assisted variable clustering\footnote{The authors' names are sorted alphabetically.
     }\hspace{.2cm}\\}

\author[1]{Dong Li}
\author[2,3]{Xinghao Qiao}
\author[1]{Cheng Yu}
\affil[1]{\it \small  Department of Statistics and Data Science, Tsinghua University, Beijing, China
}
\affil[2]{\it Faculty of Business and Economics, The University of Hong Kong, Hong Kong, China}
\affil[3]{\it Department of Statistics, London School of Economics, London, U.K.}

\date{}

  \maketitle
} \fi

\if0\blind
{
  \bigskip
  \bigskip
  \bigskip
  \begin{center}
    {\LARGE\bf Large covariance matrix estimation with factor-assisted variable clustering}
\end{center}
  \medskip
} \fi

%\maketitle

\bigskip

\spacingset{1.5}

\begin{abstract}
This paper studies the covariance matrix estimation for high-dimensional time series within a new framework that combines low-rank factor and latent variable-specific cluster structures. The popular methods based on assuming the sparse error covariance matrix after taking out common factors may be invalid for many financial applications. Our formulation postulates a latent model-based error cluster structure after removing observable factors, which not only leads to more interpretable cluster patterns but also accounts for non-sparse cross-sectional correlations among the variable-specific residuals. Our method begins with using least-squares to estimate the factor loadings, followed by identifying the latent cluster structure by thresholding the scaled covariance difference measures of residuals. A novel ratio-based criterion is introduced to determine the threshold parameter when performing the developed clustering algorithm. We then establish the cluster recovery consistency of our method and derive the convergence rates of our proposed covariance matrix estimators under different norms. Finally, we demonstrate the superior finite sample performance of our proposal over the competing methods through both extensive simulations and a real data application on minimum variance portfolio.
\end{abstract}

\bigskip \bigskip
\noindent%

\newpage
\spacingset{1.59} % DON'T change the spacing!
%\onehalfspacing

\section{Introduction}
\label{sec:intro}
The problem of estimating a covariance matrix and its inverse (precision matrix) plays a fundamental role in high-dimensional statistics and econometrics. Despite being interesting themselves, their estimation is useful in a variety of applications, such as 
portfolio allocation \citep{Engle2019Large}, market efficiency test \citep{feng2022high},
%undirected graphical models \citep{fan2016innovated}, discriminant analysis for classification \citep{fan2012road}, among others. 
partial correlation networks \citep{chang2018confidence}, among others.
In the high-dimensional setting, where the dimension $p$ is comparable to, or even greater than, the sample size $T,$ the sample covariance matrix becomes singular and performs poorly. The literature has mainly focused on imposing certain structural assumptions to develop regularization techniques for consistent estimation. Under the commonly-used sparsity assumption, various thresholding methods have been introduced; see, e.g., \cite{bickel2008covariance}, \cite{cai2011adaptive}, \cite{chen2016dynamic}, and \cite{chen2019}.

Such sparsity assumption, however, is unrealistic for many datasets in economics and finance, where variables typically exhibit high correlations. For example, financial returns in stock market are influenced by the common market factors, resulting in highly correlated variables. To relax the sparsity assumption for estimating the covariance matrix in these applications, the approximate factor model \cite[]{chamberlain1983arbitrage} is often employed:
\begin{equation}
\label{fm1}
y_{it} = \bbb_i^\T \bbf_t + u_{it}, ~~i=1, \dots,p,~t=1, \dots, T,
\end{equation}
where $y_{it}$ is the observed time series for the $i$-th variable, $\bbb_i^\T \bbf_t$ is the common component and is uncorrelated %with the idiosyncratic component $u_{it}.$
with the remaining component $u_{it}.$
The co-movement of $p$ time series $y_{1t}, \dots, y_{pt}$ is driven by $r$-vector of common factors $\bbf_t$'s with corresponding factor loadings $\bbb_i$'s. We refer to $\bSigma=(\Sigma_{ij})_{p \times p}$ and $\bSigma_u=(\Sigma_{u,ij})_{p \times p}$ as the covariance matrices of $\by_t=(y_{1t}, \dots, y_{pt})^\T$ and $\bu_t=(u_{1t}, \dots, u_{pt})^\T,$ respectively. 

Instead of directly imposing sparsity on $\bSigma,$ \cite{fan2011high} assumed that $\bSigma_u$ is sparse and applied thresholding to the residual covariance matrix after taking out the observable factors. This low-rank plus sparse covariance matrix model is extended to the case where $\bbf_t$'s are unobservable and $\bSigma_u$ is approximately sparse (i.e., $\bu_t$ is weakly cross-sectionally correlated)
in \cite{fan2013large}. They introduce an estimator for $\bSigma$ by thresholding principal orthogonal complements. See also \cite{fan2019robust}, \cite{wang2021nonparametric}, and \cite{choi2023large} along this line of research. However, the sparsity assumption imposed on $\bSigma_u$ may still be restrictive for many financial applications. For example, we estimated the measure of the sparsity level for $\bSigma_u$ (see its definition in (\ref{m_p}) and results in Figure~\ref{fig:m_p/p} of Appendix~\ref{ap.sparse}) and conducted large-scale multiple testing \cite[]{cai2016large} based on the residual covariance matrix after removing common factors, using the stock market data analyzed in \cite{zhang2023factor}, as detailed in Appendix~\ref{ap.sparse}. Our findings provide compelling empirical evidence against the sparsity assumption. Thus, the low-rank plus sparsity representation may not be appropriate for modelling the large covariance matrix in financial data.

In this paper, we propose a novel procedure for estimating $\bSigma$ by leveraging a latent cluster structure to capture the cross-sectional correlations after removing common factors. Specifically, we use the variable clustering model \cite[]{bunea2020model} to group similar individuals of $\bu_t$ 
by decomposing each individual into two uncorrelated latent components, one cluster-specific and the other representing individual fluctuation:
\begin{equation}
\label{cm1}
 u_{it} = z_{g(i), t} + e_{it},~~i=1, \dots, p, ~g(i)=1, \dots, K, 
\end{equation}
where $z_{g(i),t}$ is the mean-zero latent cluster variable at time $t$ for individual $i$ belonging to cluster $g(i),$ $K$ is the number of clusters, and $\be_t=(e_{1t}, \dots, e_{pt})^\T$ is mean-zero and cross-sectionally uncorrelated, leading to a diagonal covariance matrix $\bSigma_e.$ %of $\be_t.$
Our proposal begins with estimating the factor loadings in (\ref{fm1}) by the least squares method. We then employ the covariance distance measure along with a novel ratio-based criterion to select the thresholding parameter, thereby identifying the number of clusters $K$ and latent cluster structure in (\ref{cm1}). We finally estimate the diagonals of $\bSigma_e,$ which allows us to construct an estimator for $\bSigma$ by aggregating estimated covariance matrices for the common, cluster-specific, and individual fluctuation components.
It is noteworthy that our paper considers observable factors, as in \cite{fan2008high}, \cite{fan2011high}, \cite{fama2015five} and many other empirical applications. Unknown factors not only make it difficult to ensure the identifiability of (\ref{fm1}) when $u_{it}$ follows (\ref{cm1}), but they are also significantly outperformed by known factors in minimum variance portfolio allocation, as empirically evidenced in Section~\ref{sec.real}.

The main contributions of our paper are threefold.
\begin{itemize}
\item First, our proposal is developed within a novel framework that combines low-rank and model-based cluster structures. Unlike the existing literature that assumes sparsity in $\bSigma_u,$ our formulation adopts model (\ref{cm1}) to generate more interpretable clusters in $\bu_t$ and account for the corresponding cross-sectional dependence with a non-sparse $\bSigma_u.$ It is worth mentioning that different cluster-based factor models were also developed in \cite{ando2017clustering} and \cite{zhang2023factor}, both of which focus on discovering common and cluster-specific factors. By comparison, our paper primarily focuses on the covariance matrix estimation, which has been empirically shown to substantially outperform the covariance matrix estimators based on their models.

\item Second, though our cluster structure identification is motivated from the idea of thresholding the covariance difference measure \cite[]{bunea2020model}, its adaptation to estimate model (\ref{cm1}) and relevant covariance matrices introduces methodological innovations and additional theoretical complexities caused by the serial dependence and estimated values of $u_{it}$'s instead of true ones. While \cite{bunea2020model} applied a data-splitting method that may lack stability and efficiency for selecting the threshold parameter for independent data when performing the clustering algorithm, we develop a fast-to-implement ratio-based criterion that utilizes the data more effectively and is also valid for time series. Theoretically, we establish the cluster recovery consistency based on such criterion despite the extra errors arisen from the factor model estimation, which are not involved in \cite{bunea2020model}. Additionally, we investigate the effect of estimating unknown quantities in (\ref{fm1}) and (\ref{cm1}) on the covariance estimation by deriving the convergence rates of our proposed covariance and precision matrix estimators. 

\item Third, empirically, we not only provide interpretable clusters that facilitate the estimation of covariance and precision matrices, but also demonstrate significant outperformance compared to an extensive set of competing methods in terms of achieving lower risks and higher out-of-sample returns through a real-world application to minimum variance portfolio allocation in Section~\ref{sec.real}. This provides a useful toolkit with enhanced interpretability for practitioners in portfolio allocation and risk management.
\end{itemize}

The remainder of the paper is organized as follows. Section \ref{sec:method} presents the model setting and develops the procedure for estimating the covariance and precision matrices. 
In Section~\ref{sec.theory}, we investigate the theoretical properties of our estimators. We demonstrate the superior finite-sample performance of our proposed method through extensive simulations in Section \ref{sec.sim} and a real data application on portfolio allocation in Section~\ref{sec.real}. Section \ref{sec:discussion} concludes the paper by discussing some possible extensions. 
All technical proofs are relegated to the appendix and the supplementary material.

Throughout the paper, for any vector $\mathbf{v} = (v_1, \dots, v_p)^{\T},$ we denote its $\ell_q$-norm by 
$\|\mathbf{v}\|_q = (\sum_{i=1}^p |v_i|^q)^{1/q}$ for $1\leq q\leq \infty$. Particularly, for simplicity, write $\ell_2$-norm as $\|\mathbf{v}\|$
and $\ell_{\infty}$-norm  as $\|\mathbf{v}\|_{\max}= \max_{i}|v_{i}|,$ respectively. For any matrix $\mathbf{M} = (M_{ij})_{p\times q}$, we denote its %{\color{cyan}$\ell_1$}, 
operator, Frobenius, and elementwise maximum  %$\ell_{\infty}$ 
norms by  $\|\mathbf{M} \| = \lambda_{\max}^{1/2}(\mathbf{M}^{\T}\mathbf{M})$, $\|\mathbf{M}\|_{\tF} = (\sum_{i,j}M_{ij}^2)^{1/2}$, and $\|\mathbf{M} \|_{\max} = \max_{i,j} |M_{ij}|$, respectively, where $\lambda_{\max}(\bA)$ and $\lambda_{\min}(\bA)$ are the largest and smallest eigenvalues of a symmetric matrix $\bA$. 
%is a $p \times 1$ vector, the maximum norm and $L^2$ norm are denoted as $\|\mathbf{v}\|_{\max} = \max_{i}|v_{i}|$ and $\|\mathbf{v}\| = \sqrt{\sum_{i=1}^p v_i^2}$, respectively.
For two positive sequences $\{a_n\}$ and $\{b_n\}$, we write $a_n = O(b_n)$ if there exists a positive constant $c$ such that $a_n / b_n \le c$ and $a_n = o(b_n)$ if $a_n / b_n \rightarrow 0$. %The latter can also be written as $a_n\ll b_n$. 
We write $a_n \asymp b_n$ if $a_n = O(b_n)$ and $b_n = O(a_n)$ hold simultaneously. 
For a positive integer $p$, we write $[p] = \{1, \dots, p\}.$  Denote the sets $\mathbb{N}:=\{0, 1, 2,\dots\}$ and 
$\mathbb{N}_+:=\{1, 2,\dots\}.$ %be nonnegative and positive integer sets, respectively.
For any positive real number $x$, we use $\lceil x \rceil$ to denote the smallest integer greater than or equal to $x$. We use the notation $i \sim j$ when indices $i$ and $j$ are from the same cluster.

\section{Methodology}
\label{sec:method}

\subsection{Model setting}
\label{sec:model}
Our model jointly captures the low-rank structure (\ref{fm1}) with observable factors and the latent cluster structure (\ref{cm1}). It can be written in the vector form as
\begin{align}\label{Model2}
\begin{split}
    \mathbf{y}_t = \mathbf{B}\mathbf{f}_t + \mathbf{u}_t,\\
    \mathbf{u}_t = \bA\mathbf{z}_t + \mathbf{e}_t,
\end{split}
\end{align}
where $\mathbf{B} = (\mathbf{b}_1, \dots, \mathbf{b}_p)^{\T}$ is the $p \times r$ factor loading matrix,  
$\bz_t = (z_{1t}, \dots, z_{Kt})^{\T}$ is the $K$-vector of cluster variables, $\mathbf{e}_t = (e_{1t}, \dots, e_{pt})^{\T}$ is formed by individual fluctuations, and $\bA$ is the $p \times K$  membership matrix with ${A}_{ik} = 1$ if index $i$ belongs to latent cluster $k$ and $0$ otherwise. As a result, $\bA^{\T}\bA = \text{diag}(p_1, \dots, p_K)$, where $p_k$ is the number of variables in cluster $k \in [K]$ and $\sum_{k=1}^K p_k = p$. We will treat $r$ as fixed, independent of $(p,T),$ and allow $K$ to diverge as $p \rightarrow \infty,$ as evidenced in real data analysis.
Under our model (\ref{Model2}), the membership matrix $\bA$ corresponds to the partition of the $p$-vector of latent time series $\mathbf{u}_t$ into $K$ different clusters, denoted as $\mathcal{G} = \{G_1, \dots, G_K \}$, where the $k$-th cluster corresponds to $G_k = \{i: A_{ik} = 1 \}$ for $k \in [K].$

In Sections~\ref{sec:est} and \ref{sec:threshold}, we will present the estimation of
the membership matrix $\bA$ (i.e., the partition $\cal G$), the number of clusters $K,$ and the covariance matrix $\bSigma$ with its inverse $\bSigma^{-1}.$
%In this paper, the parameter of interest is the $p \times p$ covariance matrix, $\bSigma$, and its inverse $\bSigma^{-1}$, of $\mathbf{y}_t$, and the $p \times K$ membership matrix $\bA$. 
Throughout, we assume that the observable factors $\mathbf{f}_t$ and latent cluster variables $\mathbf{z}_t$ are uncorrelated. Then the covariance matrix of $\by_t$ from model \eqref{Model2} can be represented as
\begin{align}\label{Cov1}
\begin{split}
    \bSigma = \mathbf{B}\bSigma_{f} \mathbf{B}^{\T} + \bSigma_u,\\
    \bSigma_u = \bA\bSigma_z \bA^{\T} + \bSigma_e,
\end{split}
\end{align}
where $\bSigma_{f}$ and $\bSigma_z$ are respectively the covariance matrices of $\mathbf{f}_t$ and $\mathbf{z}_t,$ and $\bSigma_e$ is a diagonal matrix. Here we assume that the process $(\bbf_t, \bz_t, \bu_t)$ is stationary so that $\bSigma_{f}$, $\bSigma_z$, and $\bSigma_e$ do not change over time.

\begin{remark}
Along the research line of `low-rank plus sparse' modelling, it is standard to apply thresholding methods to the entries of residual covariance matrix, which places a restrictive sparsity assumption in $\bSigma_u$ (as evidenced in Appendix~\ref{ap.sparse}) and introduces non-negligible bias. In contrast, our model~(\ref{Model2}) assumes that the $p$ time series in $\by_t$ are jointly driven by both $r$ observable factors in $\bbf_t$ and $K$ latent cluster variables in $\bz_t$ contaminated with uncorrelated individual fluctuations in $\be_t.$ Consequently, $\bSigma_u$ is represented as `low-rank plus diagonal' and is not sparse. 
\end{remark}

\begin{remark}
It is worth noting that \cite{zhang2023factor} introduced a two-layer factor model with common strong factors and cluster-specific weak factors. However, the covariance estimation based on their model is not recommendable, as the gain from incorporating cluster-specific factors is often offset by the errors in estimating too many parameters in factor loadings. 
In contrast, our covariance representation \eqref{Cov1} utilizes a membership matrix $\bA$ to capture the latent cluster structure, whose estimation is determined by a single threshold parameter (see Sections~\ref{sec:est} and \ref{sec:threshold}). This simpler structure mitigates the accumulation of errors in the estimation of $\bA\bz_t,$ resulting in improved covariance estimation performance, as demonstrated by our real data analysis in Section \ref{sec.real}.
\end{remark}

\begin{remark}
\label{rmk.pcacluster}
An alternative approach is to assume unknown factors $\bbf_t$'s in (\ref{Model2}) and perform PCA on 
the sample estimator of $\bSigma,$ 
followed by applying the variable clustering step (i.e., Step~2) in Section~\ref{sec:est}  %in Section~\ref{sec.real} 
to the residual covariance matrix after removing the identified common factors. However, this method lacks theoretical guarantees, as the covariance decomposition $\bSigma = \mathbf{B}\bSigma_{f} \mathbf{B}^{\T} + \bSigma_u$ can not be identified asymptotically as $p \rightarrow \infty.$ Specifically, under a common scenario where the factors are pervasive and $p_k \asymp p/K$ uniformly over $k \in [K],$ both the first $r$ eigenvalues of $\bB\bSigma_{f}\bB^\T$ and the largest eigenvalue of $\bSigma_u$ diverge at the same rate $O(p).$  Empirically, we demonstrate in Section~\ref{sec.real} that such method of PCA with variable clustering leads to substantially inferior performance compared to our proposal. Therefore, our paper considers known factors to facilitate the subsequent estimation procedure.
\end{remark}

\subsection{Estimation procedure}
\label{sec:est}

In this section, we develop the following three-step procedure to estimate our model~(\ref{Model2}), and based on which obtain the estimated covariance and precision matrices.

\par \textbf{Step 1.}  
With the observable factors, we estimate the factor loading matrix $\bB$ by the least squares method,
$\widehat{\mathbf{B}} = (\widehat{\mathbf{b}}_1, \dots, \widehat{\mathbf{b}}_p)^{\T}$, where
\begin{equation*}
    \widehat{\mathbf{b}}_i = \arg\min_{\mathbf{b}_i} \frac{1}{T}\sum_{t=1}^T (y_{it} - \mathbf{b}_i^{\T}\mathbf{f}_t)^2, \quad i \in [p].
\end{equation*}
The covariance matrix $\bSigma_{f}$ can be estimated by its sample version
\begin{equation*}
\widehat{\bSigma}_f = \frac{1}{T-1}\sum_{t=1}^{T}(\mathbf{f}_t - \bar{\mathbf{f}})(\mathbf{f}_t - \bar{\mathbf{f}})^{\T}~~ \text{with}~~ \bar{\mathbf{f}} = \frac{1}{T}\sum_{t=1}^{T}\mathbf{f}_t.
\end{equation*}

\par \textbf{Step 2.}  
We then obtain the $p$-vector of residuals $\widehat{\mathbf{u}}_t = (\widehat{u}_{1t}, \dots, \widehat{u}_{pt})^{\T}$ with $\widehat{u}_{it} = y_{it} - \widehat{\mathbf{b}}_i^{\T} \mathbf{f}_t$ and thus construct the residual covariance matrix as %$\bSigma_u$ is 
\begin{equation}\label{Eq_SigmaU_Sample}
\widecheck{\bSigma}_u = \frac{1}{T}\sum_{t=1}^{T}\widehat{\mathbf{u}}_t \widehat{\mathbf{u}}_t^{\T} = \big(\widecheck \Sigma_{u,ij}\big)_{p \times p}.
\end{equation}
Motivated by \cite{bunea2020model}, we make use of the {\bf CO}variance {\bf D}ifference (COD) measure between two time series $\{u_{it}\}$ and $\{u_{jt}\}$ for $i\not =j,$ defined as
\begin{equation}\label{COD}
    \text{COD}(i,j) = \max_{l \not= i,j}\big|\operatorname{cov}(u_{it} - u_{jt}, u_{lt})\big|  =  \max_{l \not= i,j}\big|\Sigma_{u,il} - \Sigma_{u,jl}\big|.
\end{equation}
Under model (\ref{cm1}), some simple calculations yield that two indices $i$ and $j$ belong to the same cluster $G_k$
%different time series $\{u_{it}\}$ and $\{u_{jt}\}$ belong to the same cluster (
(i.e., driven by the same cluster-specific variable time series $\{z_{kt}\}$) if and only if $\text{COD}(i,j) = 0$. To avoid scaling issues when comparing variables with different units, we will develop the method based on the following scaled COD (sCOD): 
\begin{equation}\label{sCOD}
    \text{sCOD}(i,j) = \max_{l \not = i,j}\frac{\big|\text{cov}(u_{it} - u_{jt}, u_{lt})\big|}{\sqrt{\text{var}(u_{it} - u_{jt})\text{var}(u_{lt})} } = \max_{l \not = i,j}\big|\text{cor}(u_{it} - u_{jt}, u_{lt}) \big|,
\end{equation}
which satisfies the scale-invariance property. 

In practice, it is natural to group indices $i$ and $j$ into the same cluster if the estimator $\widehat{\text{sCOD}}(i,j)$ is below a certain threshold value $\gamma_T > 0$, where 
\begin{align}\label{sCOD_hat}
    \widehat{\text{sCOD}}(i,j) = \max_{l\not = i,j}\big|\widehat{\text{cor}}(\widehat{u}_{it} - \widehat{u}_{jt}, \widehat{u}_{lt})\big|:= \max_{l \not = i,j}\left|\frac{\widecheck{\Sigma}_{u,il} - \widecheck{\Sigma}_{u,jl} }{\sqrt{\big( \widecheck{\Sigma}_{u,ii} + \widecheck{\Sigma}_{u,jj} -2\widecheck{\Sigma}_{u,ij}\big)\widecheck{\Sigma}_{u,ll}}} \right|.
\end{align}

With the inputs of threshold parameter $\gamma_T$ and the residual covariance matrix $\widecheck{\bSigma}_{u}$, we develop the sCOD-based clustering algorithm in Algorithm~\ref{Algorithm_sCOD} to output the estimated number of clusters $\widehat K$ and the estimated partition $\widehat{\mathcal{G}} = \{\widehat{G}_1, \cdots, \widehat G_{\widehat K}\}.$ The selection of $\gamma_T$ will be discussed in Section \ref{sec:threshold}. According to $\widehat{\mathcal{G}},$ we can obtain the corresponding estimated membership matrix $\widehat\bA = (\widehat A_{ik})_{p \times \widehat K}.$ Specifically, for each $k \in [\widehat K],$ we set 
\begin{equation}\label{A.est}
\widehat A_{ik}=1~~\text{if}~~i \in \widehat G_k ~~\text{and}~~0~~\text{otherwise}.
\end{equation}
Given $\widehat \bA,$ we can obtain the least-squares estimator for the $\widehat K$-vector of cluster-variables $\bz_t$ 
and the covariance matrix estimator for $\bSigma_z,$ respectively, as
$$\widehat \bz_t = \big(\widehat \bA^\T \widehat \bA\big)^{-1} \widehat \bA^\T \widehat \bu_t~~\text{and}~~\widehat{\bSigma}_z = \frac{1}{T}\sum_{t=1}^{T}\widehat{\mathbf{z}}_t\widehat{\mathbf{z}}_t^{\T}.$$

\begin{algorithm}[h]
\caption{sCOD-based Clustering Algorithm}\label{Algorithm_sCOD}
\begin{algorithmic}
\item \textbf{Input}: $\widecheck{\bSigma}_u$ and $\gamma_T > 0$
\item \textbf{Initialization:} $\widehat{\mathcal{G}} = \{\widehat{G}_1, \ldots, \widehat{G}_p\}$ with $\widehat{G}_{i} = \{i\}$ for $i \in [p]$ and $k = p$
\item \textbf{Definition:} $\widehat{\operatorname{sCOD}}_{\min}(\widehat{G}_i, \widehat{G}_j) = \min_{a \in \widehat{G}_i, b \in \widehat{G}_j} \widehat{\operatorname{sCOD}}(a,b)$
\While{True}
    \State $(\widehat{G}_i, \widehat{G}_j) = \argmin_{\widehat{G}_i, \widehat{G}_j \in \widehat{\mathcal{G}}, \, \widehat{G}_i \neq \widehat{G}_j} \widehat{\operatorname{sCOD}}_{\min}(\widehat{G}_i, \widehat{G}_j)$
    \If {$\widehat{\operatorname{sCOD}}_{\min}(\widehat{G}_i, \widehat{G}_j) \ge \gamma_T$}
        \State \textbf{break}
    \Else
        \State $\widehat{G}_{\operatorname{new}} \gets \widehat{G}_i \cup \widehat{G}_j$
        \State $\widehat{\mathcal{G}} \gets \widehat{\mathcal{G}} \setminus \{\widehat{G}_i, \widehat{G}_j\} \cup \{\widehat{G}_{\operatorname{new}}\}$
        \State $k \gets k-1$
    \EndIf
\EndWhile 
\item \textbf{Output:} the estimated number of clusters $\widehat{K} = k$ and the estimated partition $\widehat{\mathcal{G}} = \{\widehat{G}_1, \dots, \widehat G_{\widehat K}\}$
\end{algorithmic}
\end{algorithm}

\par \textbf{Step 3.}  
The estimated covariance matrix for $\bSigma_e$ is given by $$\widehat{\bSigma}_e = \operatorname{diag}\Big( \frac{1}{T}\sum_{t=1}^{T}\widehat{\mathbf{e}}_t\widehat{\mathbf{e}}_t^{\T}\Big),$$ which is 
the diagonal matrix of $T^{-1}\sum_{t=1}^{T}\widehat{\mathbf{e}}_t\widehat{\mathbf{e}}_t^\T$ with 
$\widehat{\mathbf{e}}_t = \mathbf{y}_t - \widehat{\mathbf{B}}\mathbf{f}_t - \widehat{\bA}\widehat{\mathbf{z}}_t.$

Substituting relevant terms by their estimators in the covariance decomposition \eqref{Cov1}, we have the following substitution estimators
\begin{align}\label{Cov_hat}
\begin{split}
    \widehat{\bSigma} &= \widehat{\mathbf{B}}\widehat{\bSigma}_f \widehat{\mathbf{B}}^{\T} + \widehat{\bSigma}_u,\\
    \widehat{\bSigma}_u &= \widehat{\bA}\widehat{\bSigma}_z \widehat{\bA}^{\T} + \widehat{\bSigma}_e.
\end{split}
\end{align}
By the Sherman--Morrison--Woodbury formula, we can  obtain the precision matrix estimators
\begin{equation}\label{Cov_Inv_hat}
\begin{split}
\widehat{\bSigma}^{-1} &= \widehat{\bSigma}_u^{-1} - \widehat{\bSigma}_u^{-1}\widehat{\mathbf{B}}\big(\widehat{\bSigma}_f^{-1} + \widehat{\mathbf{B}}^{\T}\widehat{\bSigma}_u^{-1}\widehat{\mathbf{B}} \big)^{-1}\widehat{\mathbf{B}}^{\T} \widehat{\bSigma}_u^{-1},\\
\widehat{\bSigma}_u^{-1} &= \widehat{\bSigma}_e^{-1} - \widehat{\bSigma}_e^{-1}\widehat{\bA}
\big( \widehat{\bSigma}_e^{-1} + \widehat{\bA}^{\T}\widehat{\bSigma}_z^{-1}\widehat{\bA} 
\big)^{-1}\widehat{\bA}^{\T} \widehat{\bSigma}_e^{-1}.
\end{split}
\end{equation}

\begin{remark}
While our methodology is developed based on the covariance-based measure \eqref{COD}, it is natural to consider incorporating the autocovariance matrices $\bSigma_u^{(h)} = \operatorname{cov}(\mathbf{u}_{t+h}, \mathbf{u}_t) = (\Sigma_{u,ij}^{(h)})_{p \times p}$ for $h \in {\mathbb N}$ to capture the serial dependence information that is the most relevant in the context of time series. Under the assumption that  $\bSigma_e^{(h)}=\operatorname{cov}(\mathbf{e}_{t+h}, \mathbf{e}_t)$ is zero or diagonal for $h \in {\mathbb N}_+,$ it is easy to verify that indices $i$ and $j$ belong to the same
cluster if and only if $\max_{l \neq i,j}|\Sigma_{u,il}^{(h)}-\Sigma_{u,jl}^{(h)}|=0$ for $h \in \mathbb N_+.$ Motivated by this, we can propose the following {\bf A}uto{\bf CO}variance-{\bf D}ifference-based (ACOD) measure
\begin{equation*}
    \operatorname{ACOD}(i,j) =  \max_{l \not= i,j} \sum_{h=0}^{H}
    \big|\Sigma_{u,il}^{(h)}-\Sigma_{u,jl}^{(h)}\big|,
\end{equation*}
where the information is accumulated across different lags from $0$ to $H$, $\bSigma_u^{(0)}=\bSigma_u,$ and $H$ is a small positive integer that is prespecified, e.g., $1 \leq H \leq 5.$  However, in practice, especially in financial applications, the signals at off-diagonals of nonzero-lagged $\bSigma_u^{(h)}$ are typically not strong enough even at lag $h=1$, which makes the ACOD-based estimation in Step~2 less accurate. In our numerical experiments, we observed that incorporating the autocovariance information did not improve the performance of estimating the latent cluster structure. Therefore, we chose not to proceed with developing our methodology based on the ACOD measure.
\end{remark}

\subsection{Selecting the threshold parameter for clustering}
\label{sec:threshold}
In this section, we introduce a data-driven method to select the threshold parameter $\gamma_T,$ which determines the latent cluster structure in $\bu_t.$ 
While the data-splitting method for choosing the threshold parameter developed in \cite{bunea2020model} suffers from increased variability and data under-utilization and only holds for independent data,
we propose a ratio-based criterion below to choose $\gamma_T,$ which not only makes more effective use of the data but also works for the serially dependent data we consider.

Notice that, if two indices $i$ and $j$ are from the same cluster (or different clusters), we would expect the empirical measure $\widehat{\text{sCOD}}(i,j)$ in \eqref{sCOD_hat} to be close to zero (or significantly above zero). 
This motivates us to propose a ratio-based criterion as follows. For any pair $(i,j) \in [p]\times[p]$ with $i < j$, we let $\widehat{D}_{ij} = \widehat{\text{sCOD}}(i,j)$ and regard indices $i$ and $j$ from different clusters if $\widehat{D}_{ij}$ takes some large value. Specifically, we rearrange $Q = p(p-1)/2$ values of $\widehat{D}_{ij} (1\le i < j \le p)$ in a descending order $\widehat{D}_{(1)} \ge \widehat{D}_{(2)} \ge \cdots \ge \widehat{D}_{(Q)}$ and compute 
\begin{equation}\label{rho_hat}
    \widehat q = \arg\max_{m \in [Q]} \frac{\widehat{D}_{(m)} + \delta_{T}}{\widehat{D}_{(m+1)} + \delta_{T}}
\end{equation}
for some $\delta_T > 0$. Then we choose the thresholding parameter as
\begin{equation}\label{gamma_hat}
    \gamma_T = \widehat{D}_{\widehat{q}}.
\end{equation}
By implementing Algorithm~\ref{Algorithm_sCOD} with $\gamma_T$ in (\ref{gamma_hat}),  we can estimate the number of clusters and the corresponding partition as $\widehat{K}$ and $\widehat{\mathcal{G}},$ respectively. 

\begin{remark}
We include a small term $\delta_T$ in (\ref{rho_hat}) to avoid the extreme case of `0/0'. 
With a suitable order of $\delta_T$, we can establish the cluster recovery consistency of our proposal; see Theorem~\ref{thm1} in Section \ref{sec.theory}. In practice, since $\delta_T$ is usually hard to be specified, we can set $\delta_T = 0$ and replace $Q$ in \eqref{rho_hat} by $c_{q} Q$ for some constant $c_q \in (0,1),$ e.g., $c_{q}=0.75;$ see \cite{lam2012factor} and \cite{ahn2013}. 
\end{remark}

\section{Theoretical properties}
\label{sec.theory}

This section presents the theoretical properties of the proposed method. 
We use the concept of $\alpha$-mixing to characterize the serial dependence with the $\alpha$-mixing coefficients defined as
\begin{equation}\label{alpha_mixing}
    \alpha(h) = \sup_{s}\sup_{A\in \mathcal{F}_{-\infty}^{s}, \,\,  B \in \mathcal{F}_{s+h}^{\infty}} \big|\eP(A\cap B) - \eP(A)\eP(B) \big|, \quad h \in {\mathbb N}_+,
\end{equation}
where $\mathcal{F}_{s}^{s'}$ is the $\sigma$-field generated by $\{(\mathbf{e}_{t}, \mathbf{z}_t, \mathbf{f}_t): s \le t \le s'\}.$
Before presenting the theoretical results, we first impose some regularity conditions.

\begin{condition}\label{cond1}
(i) $\{(\mathbf{e}_t, \mathbf{z}_t, \mathbf{f}_t)\}_{t\ge 1}$ is strictly stationary. In addition, $\E(e_{it}) = \E(z_{kt}) = \E(e_{it}z_{kt}) = \E(e_{it}f_{jt}) = \E(z_{kt}f_{jt})=0$ for any $i \in [p]$, $j \in [r]$, $k \in [K]$, and $t \in [T]$.
(ii) $\lambda_{\min}(\bSigma_z) > c_1$ for some  constant $c_1>0.$
(iii) All the diagonals of $\bSigma_e$ are bounded away from both $0$ and $\infty.$
\end{condition}

\begin{condition}\label{cond2}
(i) There exist some constants $l>2$ and $\tau>0$ such that 
$\max_{t \in [T]}\max_{i \in [p]} \eP( |e_{it}| > x) =O(x^{-2(l + \tau)}),$
$\max_{t \in [T]}\max_{k \in [K]}\eP(|z_{kt}| > x) = O(x^{-2(l + \tau)}),$ and
$\max_{t \in [T]}\max_{j \in [r]}\eP\big(|f_{jt} - \E(f_{jt})| > x\big) = O(x^{-2(l + \tau)})$ for any $x>0.$
(ii) The mixing coefficients in (\ref{alpha_mixing}) satisfy $\alpha(h) = O\big(h^{-(l-1)(l+\tau)/\tau}\big)$ for all $h \in {\mathbb N}_+.$%, where $l$ and $\tau$ are the constants in Condition \ref{cond1} (iv).
\end{condition}

Condition~\ref{cond1}(i) assumes strict stationarity and zero correlations among $\{\be_t\},$ $\{\bz_t\}$ and $\{\bbf_t\}.$
Conditions~\ref{cond1}(ii) and (iii) require both $\bSigma_z$ and $\bSigma_e$ to be well-conditioned.
Condition~\ref{cond2} ensures the polynomial-type upper bounds for the relevant tail probabilities by employing the Fuk--Nagaev-type inequalities for $\alpha$-mixing processes \cite[]{rio2017asymptotic}, which are essential for high-dimensional time series. %and \cite{chang2018principal}. 
If we further assume that
$\max_{t \in [T]}\max_{i \in [p]}\eP(|e_{it}| > x)=O(\exp(-c_2x^{d_1})),$ $\max_{t \in [T]}\max_{k \in [K]}\eP(|z_{kt}| > x) =O( \exp(-c_2x^{d_1})),$ and
$\max_{t \in [T]}\max_{j \in [r]}\eP\big(|f_{jt} - \E(f_{jt})| > x\big) = O(\exp(-c_2x^{d_1}))$ for any $x>0$ with some constants $c_2>0$ and $d_1 \in (0,2]$ (e.g., $d_1=2$ corresponds to the sub-Gaussian case), and $\alpha(h)  =O(\exp(-c_3 h^{d_2}))$ for all $h \in {\mathbb N}_+$ with some constants $c_3>0$ and $d_2 \in (0,1],$ we can apply Bernstein-type inequalities to establish relevant non-asymptotic upper bounds, allowing $p$ to diverge at some exponential rate of $T$; see \cite{chang2018principal}.

\begin{condition}\label{cond3}
    The triplet $(p,K,T)$ satisfies $p = o(T^{(l-1)/2})$ and $K = o(p)$.
\end{condition}

\begin{condition}\label{cond4}
There exists a constant $\varsigma > 0$ such that $\min_{i \not\sim j}\text{\rm sCOD}(i,j) \ge \varsigma.$
\end{condition}

\begin{condition}\label{cond5}
The number of variables $p_k$ in each cluster satisfies $p_k \asymp p/K$ for all $k \in [K].$
\end{condition}

\begin{condition}\label{cond6}
(i) All the eigenvalues of the $r \times r$ matrix $p^{-1} \bB^\T\bB$ are bounded away from $0$ as $p \rightarrow \infty.$ 
(ii) $\|\mathbf{b}_i\|_{\max}$ is bounded away from $\infty$ for any $i \in [p]$.
\end{condition}

Condition~\ref{cond3} allows for the high dimensional setting, where $p$ grows polynomially (or exponentially under stronger assumptions as discussed below Condition \ref{cond2}) as $T$ increases, depending on the tail probabilities in Condition \ref{cond2}.
The positive constant $\varsigma$ in Condition \ref{cond4} can be interpreted as the minimal signal level required to separate different clusters, which facilitates the specifications of the true number of clusters $K$ and the partition structure $\mathcal{G}$ under model~\eqref{cm1}.
Condition~\ref{cond5} assumes the same order of $p_1, \dots, p_K$ to simplify our technical analysis. In general, we can relax this condition by allowing $p_{\min} = \min\{p_1, \dots, p_K\} \rightarrow \infty$ as $p\rightarrow\infty,$ and our established rates below will depend on $p_{\min}.$ 
Condition~\ref{cond6} requires pervasive factors, which means that they can influence a non-vanishing proportion of individual time series; see \cite{fan2013large}.

To establish the cluster recovery consistency, we reformulate using an equivalent graph representation. To this end,
define $D_{ij} = \operatorname{sCOD}(i,j)$ and $q = \big|\{D_{ij} | D_{ij} > 0, 1\le i < j \le p\}\big|,$ where $|A|$ denotes the cardinality of set $A.$ Rearranging $Q = p(p-1)/2$ values of $D_{ij}$ in a non-increasing order $D_{(1)} \ge D_{(2)} \ge \cdots \ge D_{(Q)}$, we then have $D_{(m)} \ge \varsigma$ for $m \in [q]$ and $D_{(m)} = 0$ for $m \ge q + 1$. Consider a graph $(G, E)$, where the node set $G = [p]$ consists of $p$ indices, and $E = \big\{(i,j)|D_{ij} < D_{(q)}, 1\le i < j \le p\big\}$ is the edge set. % with an edge connecting nodes $i$ and $j$ if and only if indices $i$ and $j$ are from the same cluster.
Self-loop is not allowed here. 
The true partition $\mathcal{G} = \{G_1, \dots, G_K\}$ can then be identified by splitting $(G,E)$ into $K$ isolated subgraphs $(G_1, E_1), \dots, (G_K, E_K).$  
Recall that with the threshold parameter $\gamma_T$ in \eqref{gamma_hat} (i.e., ratio-based estimator $\hat q$ in (\ref{rho_hat})) and Algorithm~\ref{Algorithm_sCOD}, we can obtain the estimated partition $\widehat{\mathcal{G}} = \{\widehat{G}_1, \dots, \widehat G_{\widehat K}\}$. Specifically, we construct an estimated graph $(G, \widehat{E})$ with the edge set $\widehat{E} = \big\{(i,j) |  \widehat{D}_{ij} < \widehat D_{\hat q}, 1\le i < j \le p\big\}$, and split it into $\widehat{K}$ isolated subgraphs $(\widehat{G}_1, \widehat{E}_1), \dots, (\widehat{G}_{\widehat{K}}, \widehat{E}_{\widehat{K}})$. 
For the sake of simplicity, we denote
\begin{equation*}
\omega(p,T) = \max\big\{p^{2/l} T^{-(l-1)/l},\,\, (T^{-1} \log p)^{1/2} \big\},
\end{equation*}
where $l > 2$ is specified in Condition~\ref{cond2}.

\begin{theorem}\label{thm1}
Let Conditions~\ref{cond1}--\ref{cond4} hold, and the threshold parameter $\delta_T$ satisfies $\delta_T / \omega(p,T) \rightarrow \infty$ and $\delta_T = o(\varsigma^2 / D_{(1)})$. Then, as $T \rightarrow \infty,$ it holds that: \\
$\mathrm{(i)}$ $\eP(\widehat{K} = K) \rightarrow 1,$ and \\
$\mathrm{(ii)}$ there exists a permutation $\phi: [K]\rightarrow [K]$ such that $\eP\big(\bigcap_{k=1}^{K}\{\widehat{G}_{\phi(k)} = G_k\}| \widehat{K} = K\big) \rightarrow 1$.
\end{theorem}

Theorem~\ref{thm1} indicates the cluster recovery consistency of our proposed method, which means that, with probability tending to one, the estimated membership matrix $\widehat{\bA}$ in \eqref{A.est} equals the true matrix $\bA$ up to some column permutation. Specifically, with probability tending to one, there exists an orthogonal permutation matrix $\bP=(P_{kk'})_{K \times K}$ such that $\widehat\bA = \bA \bP,$ where, for each $(k,k') \in [K]\times[K],$ $P_{kk'}=1$ if the $k'$-th column of $\widehat\bA$ equals the $k$-th column of $\bA$ and $0$ otherwise.

Estimating $\bSigma_u$ and its inverse is crucial for many statistical learning tasks based on factor modeling, such as statistical inference for the factor loadings and testing the capital asset pricing model. 
When $\bSigma_u$ exhibits latent cluster structure in \eqref{Model2}, the condition that $\lambda_{\max}(\bSigma_u)$ is bounded, as imposed in \cite{fan2011high,fan2013large} and \cite{wang2021nonparametric}, no longer holds. Consequently, $\widehat{\bSigma}_u$ in (\ref{Cov_hat}) cannot be consistently estimated under the operator norm in the high dimensional setting.
Motivated by \cite{fan2008high}, we employ the weighted quadratic norm, defined as $\|\bM\|_{\Sigma_u} = p^{-1/2}\|\bSigma_u^{-1/2}\bM \bSigma_u^{-1/2}\|_{\text{F}},$ where $\bM \in {\mathbb R}^{p \times p}$ and the normalization factor $p^{-1/2}$ ensures $\|\bSigma_u \|_{\bSigma_u} = 1.$ The convergence of $\widehat\bSigma_u$ can then be measured in the relative errors as $\|\widehat{\bSigma}_u - \bSigma_u\|_{\bSigma_u} = p^{-1/2}\|\bSigma_u^{-1/2}\widehat{\bSigma}_u\bSigma_u^{-1/2} - \bI_{p} \|_{\text{F}}.$ Such discussion applies analogously to the covariance matrix estimator $\widehat\bSigma$ of $\bSigma;$ see (\ref{Cov_hat}).
The following theorem gives the convergence rates for $\widehat{\bSigma}_u$ under the weighted quadratic norm and elementwise maximum norm, and for $\widehat{\bSigma}_u^{-1}$ (with bounded eigenvalues) under the operator norm.

\begin{theorem}\label{thm2}
Let Conditions~\ref{cond1}--\ref{cond5} hold. Then, as $T \rightarrow \infty$, the following assertions hold:\\ 
$\mathrm{(i)}$ $\| \widehat{\bSigma}_{u} - \bSigma_{u} \|_{\bSigma_u}= O_{\p}\big(K\omega(p,T) + K^{3/2}/p\big)$;\\
$\mathrm{(ii)}$ $\|\widehat{\bSigma}_{u} - \bSigma_{u} \|_{\max}= O_{\p}(K\omega(p,T)+ K/p)$;\\
$\mathrm{(iii)}$ If further $K\omega(p,T) = o(1)$, then $\widehat{\bSigma}_u$ has a bounded inverse with probability approaching $1$, and $\| \widehat{\bSigma}_{u}^{-1} - \bSigma_{u}^{-1} \|= O_{\p}\big(K\omega(p,T) + K/p\big)$.
\end{theorem}

\begin{remark}
\label{rmk1}
(i) The presence of $K^{3/2}/p$ and $K/p$ in the established rates arises from the estimation of the unknown cluster variables in $\mathbf{z}_t$. As $p$ increases, the enhanced information about the latent cluster structure results in more accurate estimation of $\mathbf{z}_t.$
(ii) Compared to the sample estimator $\widecheck{\bSigma}_u$ in \eqref{Eq_SigmaU_Sample}, our estimator $\widehat{\bSigma}_u$ presents several advantages, despite both having the same convergence rates in $\|\cdot\|_{\max}$ and $\|\cdot\|$ as long as $p \omega(p,T) \rightarrow \infty$.
First, under the weighted quadratic norm, $\widehat{\bSigma}_u$ converges to $\bSigma_u$ in the high-dimensional setting, whereas $\widecheck{\bSigma}_u$ does not achieve this convergence. Second, $\widehat{\bSigma}_u$ is asymptotically invertible and  $\widehat{\bSigma}_u^{-1}$ converges under the operator norm, while $\widecheck{\bSigma}_u$ becomes non-invertible when $p > T.$
\end{remark}

Similar to $\widehat\bSigma_u$, we establish  in the following theorem the convergence rates for the proposed estimators $\widehat{\bSigma}$ and  $\widehat{\bSigma}^{-1}$ under respective matrix norms.

\begin{theorem}\label{thm3}
Let  Conditions~\ref{cond1}--\ref{cond6} hold. Then, as $T \rightarrow \infty,$ the following assertions hold: \\
$\mathrm{(i)}$ $\| \widehat{\bSigma} - \bSigma \|_{\bSigma}= O_{\p}\big( \sqrt{p}\,\omega(p,T)^2 + K\omega(p,T) + K^{3/2}/p\big)$;\\
$\mathrm{(ii)}$ $\| \widehat{\bSigma} - \bSigma \|_{\max}= O_{\p}(K\omega(p,T)+ K/p)$;\\
$\mathrm{(iii)}$ If further $K\omega(p,T) = o(1)$, then $\widehat{\bSigma}$ has a bounded inverse with probability approaching $1$, and  $ \| \widehat{\bSigma}^{-1} - \bSigma^{-1} \|= O_{\p}\big(K\omega(p,T) + K/p\big)$.
\end{theorem}
 
\begin{remark}
(i) Compared to Theorem \ref{thm2}(i), the additional term $\sqrt{p}\,\omega(p,T)^2$ in Theorem~\ref{thm3}(i) arises from estimating the factor loadings. The estimation of $\bSigma^{-1},$ which is essential for determining the optimal portfolio allocation (see \eqref{weight} in Section~\ref{sec.real}), achieves the same convergence rate as $\widehat \bSigma_u^{-1}$ in Theorem \ref{thm2}(iii).
Additionally, the discussion comparing $\widehat \bSigma_u$ and $\widecheck \bSigma_u$ in Remark~\ref{rmk1}(ii) can be applied analogously to our estimator $\widehat\bSigma$ and the sample version of $\bSigma.$
\\
(ii) Theorem~\ref{thm3}(ii) and (iii) play a vital role in risk management. Given a portfolio allocation vector $\mathbf{w} \in {\mathbb R}^p,$ the true and perceived variances (i.e. risk) are given by $\mathbf{w}^{\T} \bSigma \mathbf{w}$ and $\mathbf{w}^{\T} \widehat{\bSigma} \mathbf{w}$, respectively. Then the absolute error of true and perceived risks is bounded by $| \mathbf{w}^{\T} \widehat{\bSigma} \mathbf{w}  - \mathbf{w}^{\T} \bSigma \mathbf{w} | \le \| \widehat{\bSigma} - \bSigma \|_{\max} \|\mathbf{w} \|_{1}^2$, in which Theorem~\ref{thm3}(ii) on $\|\widehat{\bSigma} - \bSigma \|_{\max}$ helps to quantify the absolute error.
Additionally, Theorem~\ref{thm3}(i) can be applied to quantify the relative error, provided that
$| \mathbf{w}^{\T} \widehat{\bSigma} \mathbf{w}  / \mathbf{w}^{\T} \bSigma \mathbf{w}  - 1 | \le \|\bSigma^{-1/2}\widehat{\bSigma}\bSigma^{-1/2} - \mathbf{I}_p \|.$
\end{remark}

\section{Simulations}
\label{sec.sim}

In this section, we conduct simulations to evaluate the finite sample performance of the proposed clustering method, as well as the covariance matrix estimators $\widehat{\bSigma}_u$ and $\widehat{\bSigma}$ and their inverses.

\subsection{Simulation setup}
Motivated by the simulation setup in \cite{fan2011high}, we simulated from a modified Fama--French five-factor model \cite[]{fama2015five} in the form of
\begin{equation}
\label{ff.m}
    y_{it} = b_{i1} f_{1t} + b_{i2} f_{2t} + b_{i3} f_{3t} + b_{i4} f_{4t} + b_{i5} f_{5t} + u_{it},
\end{equation}
where $y_{it}$ is the excess return of the $i$-th stock, and the five factors $f_{1t}, \dots, f_{5t}$ correspond to (1) market premium, (2) size premium, (3) value premium, (4) profitability premium, and (5) investment premium, respectively. 

To mimic a more realistic simulation, we calibrate the model parameters based on annualized returns of $48$ portfolios and  $5$ factors, which were downloaded from Kenneth French's website.
%To estimate the parameters in the Fama--French five-factor model, we use the data on annualized returns of $48$ portfolios and the  $5$ factors, which was downloaded from Kenneth French's website %\footnote{http://mba.tuck.dartmouth.edu/pages/faculty/ken.french/data\_library.html}.  
The excess returns $\widecheck{\mathbf{y}}_t$ for $t=1, \dots, {\widecheck T}=756$ are computed for the period from June 2020 to June 2023. The calibration procedure is outlined as follows:

(a) Based on the data $\{\widecheck{\mathbf{y}}_t\}_{t=1}^{\widecheck{T}}$ with five factors $\{\widecheck{\mathbf{f}}_t\}_{t=1}^{\widecheck{T}},$ we fit the model in (\ref{ff.m}) using the least squares method, and obtain the $48 \times 5$ factor loading matrix $\widecheck{\mathbf{B}}.$ 
For the rows of $\widecheck{\mathbf{B}},$ we compute their sample mean vector $\bmu_{b}$ and sample covariance matrix $\bSigma_{b}$, which are reported in Table S.1 of the supplementary material. The factor loadings $\mathbf{b}_i = (b_{i1}, \dots, b_{i5})^{\T}$ for $i \in [p]$ are then sampled independently from ${\cal N}_5(\bmu_b, \bSigma_b)$.

(b) We consider two cases for the membership matrix $\bA=(A_{ik})_{p \times K}:$
\begin{itemize}
\item%[I.] 
{\bf Balanced case}: The number of variables in each cluster is set as $p_k = \lceil p /K \rceil$ for $k \in [K-1]$, with the final cluster size $p_K$ adjusted to ensure that the total size equals $p$, i.e., $p_K = p - \sum_{k=1}^{K-1} p_k$.
\item%[II.] 
{\bf Imbalanced case}: The sizes of the $K$ clusters, $p_1, \dots , p_K,$ are generated from a multinomial distribution $\text{Mult}(p; \pi_1, \dots, \pi_K)$, where the cluster proportions are specified as: $\pi_{1}= \dots = \pi_{\lceil K/3 \rceil} = 3\iota$, $\pi_{\lceil K/3 \rceil + 1}= \dots = \pi_{2\lceil K/3 \rceil} = 2\iota$ and $\pi_{2\lceil K/3 \rceil + 1}= \dots = \pi_{K} = \iota$, subject to $\sum_{k=1}^{K}\pi_{k} = 1$. 
\end{itemize}
%The membership matrix $\bA$ is then defined as 
We then define $A_{ik} = 1$ if $ \sum_{l=0}^{k-1} p_{l} + 1 \le i \le  \sum_{l=0}^{k} p_{l} $ for $k \in [K]$ with $p_0 = 0$,  and $A_{ik} = 0$ otherwise.

(c) We generate the error covariance matrix $\bSigma_e$ as follows.
%$$\{\mathbf{e}_t\}$ are generated as follows.  
Let $\widecheck{\mathbf{u}}_t = \widecheck{\mathbf{y}}_t - \widecheck{\mathbf{B}}\widecheck{\mathbf{f}}_t$, then $\widecheck{\bSigma}_z = \sum_{t=1}^{\widecheck{T}}\widecheck{\mathbf{z}}_t \widecheck{\mathbf{z}}_t^{\T}/\widecheck{T}$, and $\widecheck{\bSigma}_e = \text{diag}\big( \sum_{t=1}^{\widecheck{T}}\widecheck{\mathbf{e}}_t\widecheck{\mathbf{e}}_t^{\T}/\widecheck{T} \big)$ with $\widecheck{\mathbf{z}}_t = (\bA^{\T} \bA)^{-1}\bA^{\T}\widecheck{\mathbf{u}}_t $ and $\widecheck{\mathbf{e}}_t = \widecheck{\mathbf{u}}_t - \bA\widecheck{\mathbf{z}}_t$. For each portfolio $i$, we calculate the sample standard deviation $\widecheck{\sigma}_i$ of the residuals $\{\widecheck{e}_{it}\}_{t=1}^{\widecheck{T}}$. We then calculate the minimum, maximum, sample mean, and sample standard deviation of $\{\widecheck{\sigma}_i\}_{i=1}^{48}$, denoted as $\sigma_{\min}$, $\sigma_{\max}$, $\bar{\sigma}$ and $s_{\sigma}$, respectively.
We then construct the matrix $\bSigma_e = \text{diag}(\sigma_1^2, \dots, \sigma_p^2)$, where $\sigma_1, \dots, \sigma_p$ are sampled independently from a Gamma distribution $\Gamma(\alpha, \beta)$, with mean $\alpha\beta$ and standard deviation $\sqrt{\alpha}\beta$. The parameters $\alpha$ and $\beta$ are chosen to ensure $\alpha \beta = \bar{\sigma}$ and $\sqrt{\alpha} \beta = s_{\sigma}$. Note that we only keep the $\sigma_i$'s that lie between $\sigma_{\min}$ and $\sigma_{\max}$. %This process constructs the matrix $\bSigma_e$.

(d) We assume that both $\{\mathbf{f}_t\}$ and $\{\mathbf{z}_t\}$ follow the VAR(1) models, $\mathbf{f}_t = \bmu_{f} + \bPhi_{f} \mathbf{f}_{t-1} + \bepsilon_{f,t}$ and $\mathbf{z}_t =  \bPhi_z \mathbf{z}_{t-1} + \bepsilon_{z,t}$, where $\bepsilon_{f,t}$'s 
and $\bepsilon_{z,t}$'s are sampled independently from ${\cal N}_5(\mathbf{0}, \bSigma_{f,\epsilon})$ and %$\bepsilon_{z,t}$ are i.i.d from 
${\cal N}_K(\mathbf{0}, \bSigma_{z,\epsilon}),$ respectively, with $\bSigma_{f,\epsilon} = \bSigma_{f} - \bPhi_{f} \bSigma_{f} \bPhi_{f}^{\T}$ and $\bSigma_{z,\epsilon} = \bSigma_z - \bPhi_z \bSigma_z \bPhi_z^{\T}$. We estimate the parameters $\bmu_f, \bSigma_f, \bPhi_f, \bSigma_z$ and $\bPhi_z$ from the data $\{(\widecheck \by_t, \widecheck \bbf_t)\}_{t=1}^{\widecheck{T}},$ where the results are summarized in Tables S.2 and S.3 of the supplementary material.

For each triplet $(p,K,T)$, we generate $\{\mathbf{f}_t\}_{t=1}^{T}$ and $\{\mathbf{y}_t\}_{t=1}^{T}$ according to the following steps:
\begin{enumerate}[Step 1:]
    \item Sample $\{\mathbf{b}_i\}_{i=1}^{p}$ from  ${\cal N}_5(\bmu_b, \bSigma_b)$ independently, and construct $\mathbf{B} = (\mathbf{b}_1, \dots,\mathbf{b}_p)^{\T}.$
    \item Sample $\{\mathbf{e}_t\}_{t=1}^{T}$ from ${\cal N}_p(\mathbf{0}, \bSigma_e)$ independently.
    \item Generate $\{\mathbf{f}_t\}_{t=1}^{T}$ from $\mathbf{f}_t = \bmu_{f} + \bPhi_{f} \mathbf{f}_{t-1} + \bepsilon_{f,t}.$
    \item Generate $\{\mathbf{z}_t\}_{t=1}^{T}$ from $\mathbf{z}_t = \bPhi_z \mathbf{z}_{t-1} + \bepsilon_{z,t}.$
    \item Calculate $\{\mathbf{y}_t\}_{t=1}^{T}$ from $\mathbf{y}_t = \mathbf{B} \mathbf{f}_t + \bA \mathbf{z}_t + \mathbf{e}_t.$
\end{enumerate}

\subsection{Results}
We generate $T=300, 400, 500$ observations of $p=200, 300, 400$ variables and $K=6, 12$ clusters under both the balanced and imbalanced cases, and ran each simulation $100$ times. We assess the performance of the cluster recovery of our method in terms of the relative frequency estimate for $\eP(\widehat K = K)$ and the Adjusted Rand Index (ARI) \cite[]{hubert1985comparing}, whose definition can be found in Section S.2.2
of the supplementary material. The ARI takes values in $[-1,1]$ with larger values indicating higher level of similarity between the estimated and true cluster structures. We report the numerical results in Table~\ref{Table:Clustering}. As one would expect, more accurate clustering is obtained 
as $T$ or $p$ increases. When $K$ gets larger or the cluster structure becomes more heterogeneous (from balanced to imbalanced), it poses more challenges in identifying the latent cluster structure. Nevertheless, we can still achieve good cluster recovery consistency when $T$ and $p$ are sufficiently large.

\begin{table}[htbp]
  \centering
  \small
  \caption{The average relative frequencies $\widehat K= K$ and the average ARIs over 100 simulation runs.
%value of $\mathcal{R}(K)$ and ARI
}
    \begin{tabular}{cccccccccccc}
    \toprule
          &       &       & \multicolumn{4}{c}{$\eP(\widehat K= K)$}      &       & \multicolumn{4}{c}{ARI} \\
    \midrule
    \multirow{2}[4]{*}{$T$} & \multirow{2}[4]{*}{$p$} &       & \multicolumn{2}{c}{Balanced} & \multicolumn{2}{c}{Imbalanced} &       & \multicolumn{2}{c}{Balanced} & \multicolumn{2}{c}{Imbalanced} \\
\cmidrule{4-7}\cmidrule{9-12}          &       &       & $K=6$   & $K=12$  & $K=6$   & $K=12$  &       & $K=6$   & $K=12$  & $K=6$   & $K=12$ \\
    \midrule
    \multirow{3}[2]{*}{300} & 200   &       & 0.81  & 0.69  & 0.78  & 0.65  &       & 0.86  & 0.77  & 0.79  & 0.72  \\
          & 300   &       & 0.92  & 0.77  & 0.88  & 0.71  &       & 0.92  & 0.82  & 0.80  & 0.74  \\
          & 400   &       & 0.98  & 0.91  & 0.94  & 0.86  &       & 0.93  & 0.91  & 0.82  & 0.76  \\
    \midrule
    \multirow{3}[2]{*}{400} & 200   &       & 0.93  & 0.88  & 0.88  & 0.84  &       & 0.93  & 0.87  & 0.89  & 0.86  \\
          & 300   &       & 0.99  & 0.92  & 0.95  & 0.88  &       & 0.95  & 0.89  & 0.92  & 0.89  \\
          & 400   &       & 1     & 0.96  & 0.98  & 0.92  &       & 0.98  & 0.94  & 0.93  & 0.90  \\
    \midrule
    \multirow{3}[2]{*}{500} & 200   &       & 1     & 0.92  & 0.96  & 0.91  &       & 0.97  & 0.93  & 0.94  & 0.90  \\
          & 300   &       & 1     & 0.98  & 1     & 0.95  &       & 0.98  & 0.94  & 0.95  & 0.93  \\
          & 400   &       & 1     & 1     & 1     & 1     &       & 0.99  & 0.95  & 0.97  & 0.94  \\
    \bottomrule
    \end{tabular}%
  \label{Table:Clustering}%
\end{table}%

In terms of estimation accuracy, Tables \ref{Table:relative_norm_of_Sigma_hat} and 
\ref{Table:Max_norm_of_Sigma_hat} present numerical summaries of losses measured by the weighted quadratic norm and elementwise maximum norm for our proposed covariance matrix estimator in \eqref{Cov_hat} and the sample estimator of $\bSigma.$ Table~\ref{Table:L2_norm_of_Sigma_hat_inv} also reports results for losses under the operator norm for the corresponding precision matrix estimators. We observe a few apparent trends.
First, across all $(T,p,K),$ the proposed cluster-based estimators consistently outperform the sample estimators by a large margin regardless of the heterogeneity of the cluster structure. In particular, although the cluster-based estimator achieves the same convergence rate under the elementwise maximum norm as the sample estimator, it exhibits superior finite-sample performance.
Second, the accuracy of covariance matrix estimation improves as $T$ enlarges or as the cluster structure becomes more balanced with a reduced $K.$ Additionally, as $p$ increases, the weighted quadratic norm loss for the cluster-based estimator decreases, whereas the performance of the sample estimator deteriorates.
Third, in the high-dimensional setting with $p\geq T,$ while the sample estimator is not invertible, our cluster-based estimator still perform reasonable well. Similar trends can also be observed from the estimation results for $\bSigma_u$ and $\bSigma_u^{-1},$ as evidenced by Tables S.4--S.6 in Section S.2.3 of the supplementary material.

\begin{table}[htbp]
  \centering
  \small
  \caption{The averages and standard errors (in parentheses) of $\|\widehat{\bSigma} - \bSigma\|_{\bSigma}$ over 100 simulation runs.}
    \begin{tabular}{cccccccccccc}
    \toprule
         &      &      & \multicolumn{4}{c}{Balanced}  &       & \multicolumn{4}{c}{Imbalanced} \\
\cmidrule{4-12}          &       &       & \multicolumn{2}{c}{Cluster} & \multicolumn{2}{c}{Sample} &       & \multicolumn{2}{c}{Cluster} & \multicolumn{2}{c}{Sample} \\
    \midrule
    $T$     & $p$     &      & $K=6$   & $K=12$  & $K=6$   & $K=12$  &       & $K=6$   & $K=12$  & $K=6$   & $K=12$ \\
    \midrule
    \multirow{6}[2]{*}{300} & \multirow{2}[1]{*}{200} &       & 0.43  & 0.46  & 0.81  & 0.82  &       & 0.47  & 0.49  & 0.82  & 0.82  \\
          &       &       & (0.01) & (0.01) & (0.02) & (0.02) &       & (0.01) & (0.01) & (0.02) & (0.02) \\
          & \multirow{2}[0]{*}{300} &       & 0.37  & 0.39  & 0.99  & 1.00  &       & 0.39  & 0.41  & 1.00  & 1.00  \\
          &       &       & (0.01) & (0.01) & (0.02) & (0.02) &       & (0.01) & (0.01) & (0.02) & (0.02) \\
          & \multirow{2}[1]{*}{400} &       & 0.33  & 0.37  & 1.15  & 1.16  &       & 0.35  & 0.39  & 1.16  & 1.16  \\
          &       &      & (0.01) & (0.01) & (0.03) & (0.03) &       & (0.01) & (0.01) & (0.03) & (0.03) \\
    \midrule
    \multirow{6}[2]{*}{400} & \multirow{2}[1]{*}{200} &       & 0.42  & 0.43  & 0.71  & 0.71  &       & 0.42  & 0.44  & 0.71  & 0.71  \\
          &       &       & (0.01) & (0.01) & (0.02) & (0.02) &       & (0.01) & (0.01) & (0.02) & (0.02) \\
          & \multirow{2}[0]{*}{300} &       & 0.35  & 0.36  & 0.86  & 0.87  &       & 0.36  & 0.37  & 0.87  & 0.87  \\
          &       &       & (0.01) & (0.01) & (0.02) & (0.02) &       & (0.01) & (0.01) & (0.02) & (0.02) \\
          & \multirow{2}[1]{*}{400} &       & 0.31  & 0.34  & 1.00  & 1.00  &       & 0.32  & 0.35  & 1.00  & 1.00  \\
          &       &      & (0.01) & (0.01) & (0.02) & (0.02) &       & (0.01) & (0.01) & (0.02) & (0.03) \\
    \midrule
    \multirow{6}[2]{*}{500} & \multirow{2}[1]{*}{200} &       & 0.40  & 0.41  & 0.64  & 0.63  &       & 0.42  & 0.42  & 0.63  & 0.64  \\
          &       &       & (0.01) & (0.01) & (0.02) & (0.02) &       & (0.01) & (0.01) & (0.02) & (0.02) \\
          & \multirow{2}[0]{*}{300} &       & 0.32  & 0.34  & 0.77  & 0.78  &       & 0.35  & 0.37  & 0.78  & 0.78  \\
          &       &       & (0.01) & (0.01) & (0.02) & (0.02) &       & (0.01) & (0.01) & (0.02) & (0.02) \\
          & \multirow{2}[1]{*}{400} &       & 0.29  & 0.31  & 0.89  & 0.90  &       & 0.31  & 0.32  & 0.90  & 0.90  \\
          &       &      & (0.01) & (0.01) & (0.02) & (0.02) &       & (0.01) & (0.01) & (0.02) & (0.03) \\
    \bottomrule
    \end{tabular}%
\label{Table:relative_norm_of_Sigma_hat}%
\end{table}%

\begin{table}[htbp]
  \centering
  \small
  \caption{The averages and standard errors (in parentheses) of $\|\widehat{\bSigma} - \bSigma\|_{\max}$ over 100 simulation runs.}
    \begin{tabular}{cccccccccccc}
    \toprule
         &      &      & \multicolumn{4}{c}{Balanced}  &       & \multicolumn{4}{c}{Imbalanced} \\
\cmidrule{4-12}          &       &       & \multicolumn{2}{c}{Cluster} & \multicolumn{2}{c}{Sample} &       & \multicolumn{2}{c}{Cluster} & \multicolumn{2}{c}{Sample} \\
    \midrule
    $T$     & $p$     &      & $K=6$   & $K=12$  & $K=6$   & $K=12$  &       & $K=6$   & $K=12$  & $K=6$   & $K=12$ \\
    \midrule
    \multirow{6}[2]{*}{300} & \multirow{2}[1]{*}{200} &       & 0.79  & 0.81  & 1.04  & 1.11  &       & 0.86  & 0.89  & 1.12  & 1.12  \\
          &       &       & (0.02) & (0.02) & (0.03) & (0.02) &       & (0.02) & (0.02) & (0.02) & (0.02) \\
          & \multirow{2}[0]{*}{300} &       & 0.81  & 0.82  & 1.12  & 1.12  &       & 0.88  & 0.88  & 1.13  & 1.13  \\
          &       &       & (0.02) & (0.02) & (0.02) & (0.02) &       & (0.02) & (0.03) & (0.02) & (0.02) \\
          & \multirow{2}[1]{*}{400} &       & 0.86  & 0.87  & 1.12  & 1.13  &       & 0.89  & 0.92  & 1.14  & 1.13  \\
          &       &      & (0.02) & (0.02) & (0.03) & (0.03) &       & (0.03) & (0.02) & (0.03) & (0.03) \\
    \midrule
    \multirow{6}[2]{*}{400} & \multirow{2}[1]{*}{200} &       & 0.75  & 0.71  & 1.04  & 1.05  &       & 0.76  & 0.78  & 1.11  & 1.13  \\
          &       &       & (0.02) & (0.02) & (0.02) & (0.02) &       & (0.02) & (0.02) & (0.02) & (0.02) \\
          & \multirow{2}[0]{*}{300} &       & 0.76  & 0.75  & 1.08  & 1.10  &       & 0.82  & 0.82  & 1.12  & 1.15  \\
          &       &       & (0.02) & (0.02) & (0.03) & (0.02) &       & (0.02) & (0.02) & (0.03) & (0.02) \\
          & \multirow{2}[1]{*}{400} &       & 0.77  & 0.79  & 1.10  & 1.16  &       & 0.84  & 0.86  & 1.13  & 1.17  \\
          &       &      & (0.02) & (0.02) & (0.03) & (0.02) &       & (0.02) & (0.02) & (0.02) & (0.02) \\
    \midrule
    \multirow{6}[2]{*}{500} & \multirow{2}[1]{*}{200} &       & 0.66  & 0.68  & 0.96  & 0.93  &       & 0.72  & 0.74  & 1.03  & 1.01  \\
          &       &       & (0.02) & (0.02) & (0.02) & (0.02) &       & (0.02) & (0.02) & (0.02) & (0.02) \\
          & \multirow{2}[0]{*}{300} &       & 0.68  & 0.71  & 0.97  & 0.99  &       & 0.73  & 0.76  & 1.05  & 1.03  \\
          &       &       & (0.02) & (0.02) & (0.02) & (0.02) &       & (0.02) & (0.02) & (0.02) & (0.02) \\
          & \multirow{2}[1]{*}{400} &       & 0.72  & 0.74  & 1.04  & 1.04  &       & 0.76  & 0.77  & 1.06  & 1.05  \\
          &       &      & (0.02) & (0.02) & (0.03) & (0.02) &       & (0.02) & (0.02) & (0.03) & (0.02) \\
    \bottomrule
    \end{tabular}%
  \label{Table:Max_norm_of_Sigma_hat}%
\end{table}%

\begin{table}[htbp]
  \centering
  \small
  \caption{The averages and standard errors (in parentheses) of $\|\widehat{\bSigma}^{-1} - \bSigma^{-1}\|$ over 100 simulation runs.}
    \begin{tabular}{cccccccccccc}
    \toprule
     &      &      & \multicolumn{4}{c}{Balanced}  &       & \multicolumn{4}{c}{Imbalanced} \\
\cmidrule{4-12}          &       &       & \multicolumn{2}{c}{Cluster} & \multicolumn{2}{c}{Sample} &       & \multicolumn{2}{c}{Cluster} & \multicolumn{2}{c}{Sample} \\
    \midrule
    $T$     & $p$     &      & $K=6$   & $K=12$  & $K=6$   & $K=12$  &       & $K=6$   & $K=12$  & $K=6$   & $K=12$ \\
    \midrule
    \multirow{6}[2]{*}{300} & \multirow{2}[1]{*}{200} &       & 51.92  & 59.12  & 230.91  & 222.93  &       & 51.69  & 59.15  & 226.86  & 217.14  \\
          &       &       & (1.59) & (1.77) & (4.62) & (4.23) &       & (1.58) & (1.85) & (4.42) & (4.55) \\
          & \multirow{2}[0]{*}{300} &       & 76.47  & 80.96  & \multirow{2}[0]{*}{---} & \multirow{2}[0]{*}{---} &       & 92.83  & 98.00  & \multirow{2}[0]{*}{---} & \multirow{2}[0]{*}{---} \\
          &       &       & (1.89) & (1.97) &       &       &       & (1.61) & (1.92) &       &  \\
          & \multirow{2}[1]{*}{400} &       & 116.29  & 119.00  & \multirow{2}[1]{*}{---} & \multirow{2}[1]{*}{---} &       & 124.59  & 132.64  & \multirow{2}[1]{*}{---} & \multirow{2}[1]{*}{---} \\
          &       &      & (1.94) & (2.02) &       &       &       & (1.82) & (2.12) &       &  \\
    \midrule
    \multirow{6}[2]{*}{400} & \multirow{2}[1]{*}{200} &       & 51.74  & 54.15  & 101.37  & 102.10  &       & 52.75  & 54.18  & 106.24  & 104.39  \\
          &       &       & (1.69) & (1.65) & (3.63) & (3.85) &       & (1.77) & (1.88) & (3.78) & (3.77) \\
          & \multirow{2}[0]{*}{300} &       & 79.57  & 71.41  & 481.91  & 497.13  &       & 80.15  & 81.09  & 496.75  & 520.20  \\
          &       &       & (1.75) & (1.83) & (10.94) & (11.22) &       & (1.93) & (2.07) & (10.69) & (11.03) \\
          & \multirow{2}[1]{*}{400} &       & 104.44  & 108.48  & \multirow{2}[1]{*}{---} & \multirow{2}[1]{*}{---} &       & 116.04  & 120.91  & \multirow{2}[1]{*}{---} & \multirow{2}[1]{*}{---} \\
          &       &      & (1.89) & (1.82) &       &       &       & (2.03) & (1.94) &       &  \\
    \midrule
    \multirow{6}[2]{*}{500} & \multirow{2}[1]{*}{200} &       & 47.41  & 49.37  & 79.15  & 81.70  &       & 48.94  & 49.70  & 83.03  & 83.38  \\
          &       &       & (1.56) & (1.48) & (3.17) & (3.44) &       & (1.74) & (1.81) & (3.40) & (3.45) \\
          & \multirow{2}[0]{*}{300} &       & 60.40  & 63.38  & 189.80  & 196.37  &       & 66.84  & 67.46  & 199.58  & 207.45  \\
          &       &       & (1.68) & (1.58) & (9.05) & (9.52) &       & (2.05) & (2.09) & (10.54) & (10.14) \\
          & \multirow{2}[1]{*}{400} &       & 88.60  & 98.07  & 845.48  & 836.62  &       & 97.56  & 102.70  & 861.63  & 874.36  \\
          &       &      & (1.74) & (1.73) & (20.78) & (20.11) &       & (2.11) & (1.91) & (20.94) & (20.63) \\
    \bottomrule
    \end{tabular}%
  \label{Table:L2_norm_of_Sigma_hat_inv}%
\end{table}%

\newpage
\section{Real data analysis}
\label{sec.real}

In this section, we illustrate our developed methodology using the daily returns of the stocks listed in S$\&$P 500 from January, 2015, to December, 2019, as studied in \cite{zhang2023factor}. We removed stocks that were not traded on every trading day during this period, resulting in a total of $p = 477$ stocks, each traded over $T=1259$ days. Denote the $p$-vector of stock daily returns by $\by_t$ for $t \in [T].$
The stocks are from $11$ industry sectors: Communication Services (CM), Consumer Discretionary (CD), Consumer Staples (CS), Energy (EN), Financials (FI), Health Care (HC), Industrials (IN), Information Technology (IT), Materials (MA), Real Estate (RE), and Utilities (UT). 

\subsection{Clustering result}
\begin{figure}[h]
    \centering
    \begin{subfigure}
        \centering
        \includegraphics[width=8.5cm]{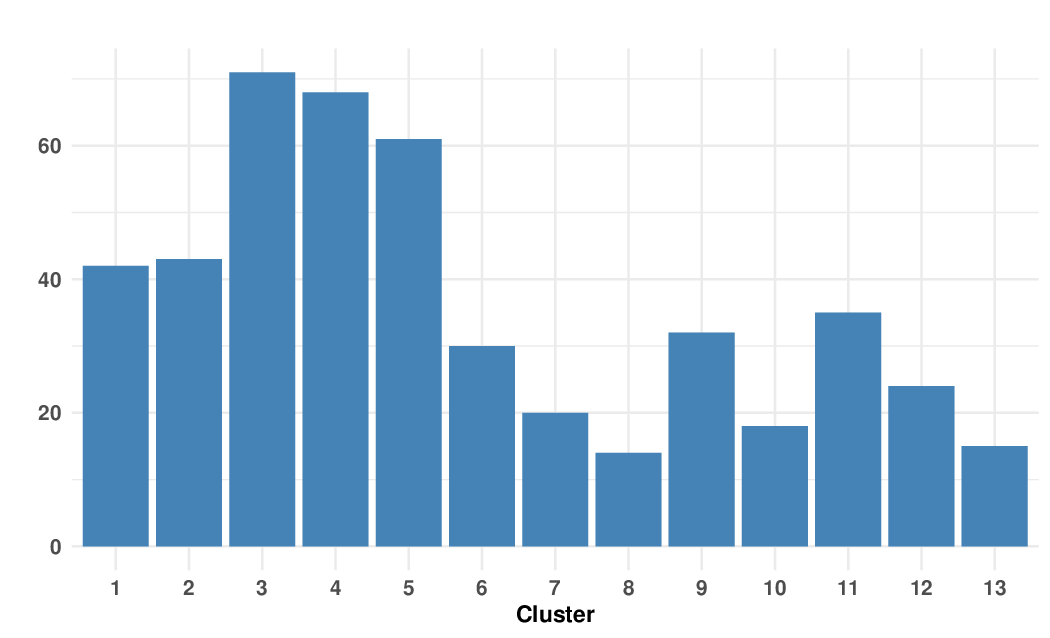}
        %\caption{Number of stocks in each cluster}
        %\label{fig:Cluster_count}
    \end{subfigure}
    %\vspace{1cm} % Adjust space between the figures if needed
    \begin{subfigure}
        \centering
        \includegraphics[width=8.5cm]{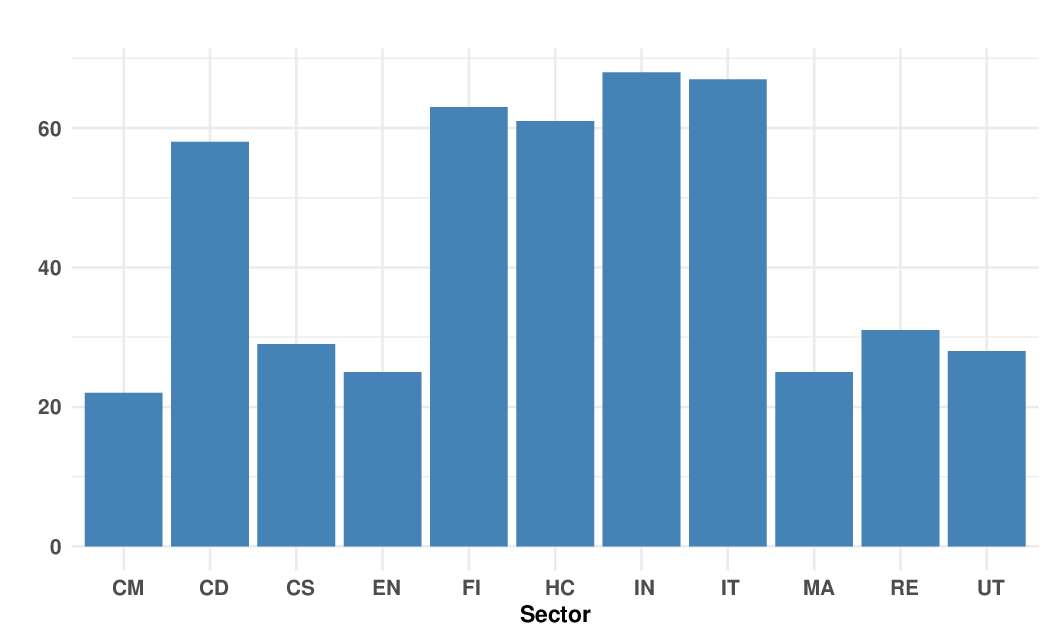}
        %\caption{Number of stocks in each sector}
        %\label{fig:Sector_count}
    \end{subfigure}
    \caption{Number of stocks in each identified cluster (left) and each sector (right).}
    \label{fig:Stock_count}
\end{figure}

To explore the latent cluster structure, we implemented the three-step procedure in Section \ref{sec:est} using 
$r=5$ observable factors in the Fama–French five-factor model in (\ref{ff.m}).
%\cite[]{fama2015five}. 
We adopt the ratio-based method in Section~\ref{sec:threshold} to estimate the number of latent clusters as $\widehat K=13.$ 
Figure \ref{fig:Stock_count} presents two bar charts, respectively depicting the number of stocks in each identified cluster and each sector. These plots indicate an imbalanced distribution of stocks across different clusters or sectors.

\begin{figure}[htbp]
	\centering
	\includegraphics[height=8cm, width=10cm]{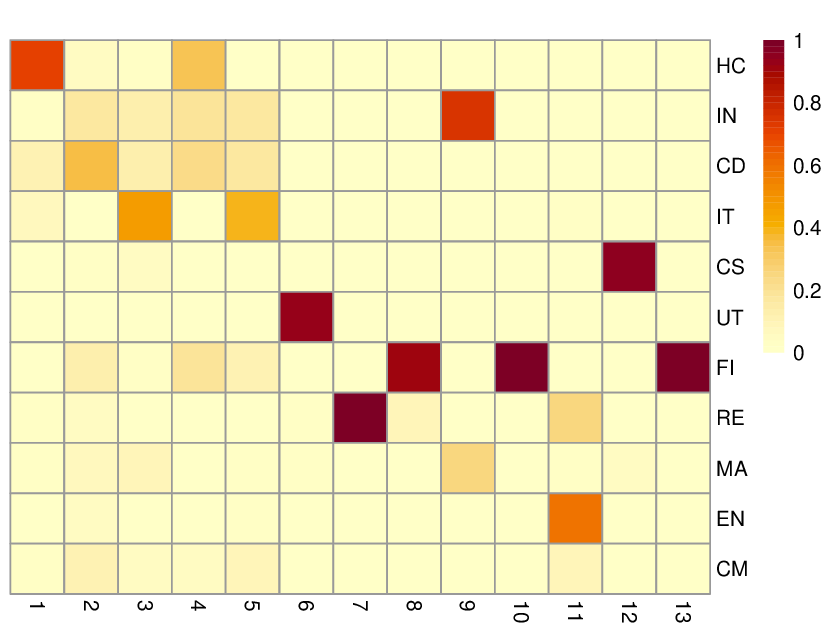}
	\caption{Heatmaps of the distributions of the stocks in each of $\widehat K=13$ identified clusters over 11 sectors.}
	\label{cluster.result}
\end{figure}

To present the identified clusters, we define $11 \times \widehat K$ matrix with $p_{ij}/p_j \in [0,1]$ as its $(i,j)$-th entry, where all $p$ stocks are from 11 sectors, $p_j$ is the number of stocks in the $j$-th cluster, $p_{ij}$ is the number of stocks in the $i$-th sector that are assigned to the $j$-th cluster and $\sum_{i=1}^{11} p_{ij}/p_j = 1$ for $j \in [\widehat K].$ %$and $\sum_{=1}^{11} n_i=p.$ 
Figure \ref{cluster.result} plots the heatmap of this $11 \times \widehat{K}$ matrix. Several interpretable patterns are observed.
Cluster 1 predominantly includes companies in the Healthcare sector.
Cluster 2 mainly consists of companies in the Consumer Discretionary and Industrial sectors.
Cluster 3 is primarily composed of companies from Information Technology, along with some from other sectors.
Cluster 4 is a mix of companies from the Healthcare, Industrials, Consumer Discretionary, and Financials sectors.
Cluster 5 primarily contains companies in Information Technology, along with some companies from Industrials and Consumer Discretionary.
Clusters 6 and 12 each contain almost all companies from Utilities and Consumer Staples, respectively.
Cluster 7 only comprises the majority of companies from the Real Estate sector.
Clusters 8, 10 and 13 each predominately contains a number of companies in Financials.
Cluster 9 mainly includes companies from the Industrials and Materials sectors. Cluster 11 contains the majority of companies in the Energy sector, along with a few from the Real Estate sector.

While both our study and \cite{zhang2023factor} have discovered interactive patterns between pre-specified sectors and identified clusters, our study yields more interpretable results for the latent cluster structure within the S$\&$P 500. For instance, we identify a unique Cluster 12 for Consumer Staples, rather than grouping it with other sectors. 
This reflects its unique market behavior, providing a more precise representation of this sector.
Moreover, our results of dividing Financials into several distinct clusters offer a finer partition of this sector, which more accurately captures variations that may not be evident in a single cluster. These findings provide valuable insights into sector-specific behaviors and interactions.

\subsection{Minimum variance portfolio}
We next apply our proposed covariance matrix estimation to construct minimum variance portfolios under two weighting schemes: one unconstrained and one with short-sales constrains. In the unconstrained case, which allows for short-sales, we obtain the optimal portfolio allocation by minimizing the empirical risk of the portfolio:
\begin{equation}\label{weight}
\min_{\bw \in {\mathbb R}^p} \bw^{\T} \widetilde\bSigma \bw, ~~\text{subject to}~~\bw^{\T}\mathbf{1}_p = 1,
\end{equation}
where $\widetilde\bSigma$ is an estimator for $\bSigma$ and $\mathbf{1}_p$ is the $p$-vector of ones. The analytical solution to (\ref{weight}) is
%\begin{equation*}
$\widehat\bw_1 = \widetilde\bSigma^{-1}\mathbf{1}_p / \big(\mathbf{1}_p^{\T} \widetilde\bSigma^{-1}\mathbf{1}_p\big).$
%\end{equation*}
In the presence of short-sales constrains, we add non-negativity constraints on the weights $w_1, \dots, w_p$ to the optimization problem (\ref{weight}). The corresponding optimal portfolio allocation $\widehat\bw_2$ can then be determined using numerical optimization techniques.

To evaluate the performance of the constructed portfolio allocation $\widehat\bw$ on the test set with size $T_{\text{test}}$, we use the following three performance measures adopted in \cite{Engle2019Large}, i.e., out-of-sample averaged return (AV), standard deviation of return (SD), and information ratio (IR), respectively, defined as:
\begin{equation}
\label{three.metric}
\operatorname{AV} = \frac{1}{T_{\text{test}}}\sum_{t=1}^{T_{\text{test}}}\widehat\bw^{\T}\widetilde\by_{t}, 
\quad
\operatorname{SD} = \Big\{\frac{1}{T_{\text{test}}-1}\sum_{t=1}^{T_{\text{test}}}(\widehat\bw^{\T}\widetilde\by_{t} - \operatorname{AV})^2\Big\}^{1/2},
\quad
\operatorname{IR} = \frac{\operatorname{AV}}{\operatorname{SD}}, 
\end{equation}
where $\widetilde\by_t$ is the $p$-vector of daily returns on the $t$-th trading day in the test set. We employ a rolling window method. 
Specifically, on each trading day of the test set from January 2017 to December 2019, we estimate $\bSigma$ using different methods based on the training set consisting of $T_{\text{train}}=504$ trading days from the preceding 2 years. We then solve both the unconstrained and constrained optimization problems related to (\ref{weight}), which allows us to obtain the corresponding optimal portfolio allocations $\widehat\bw_1$ and $\widehat\bw_2.$ 
% At the end of the month after $21$ trading days, we calculate the three annualized evaluation metrics in (\ref{three.metric}).
%At the end of the whole testing period, 
Finally, we calculate the three performance measures in (\ref{three.metric}), which are annualized (i.e., AV is multiped by $252$ and SD is multiplied by $\sqrt{252}$) and in percent to facilitate interpretation.

We compare our proposed observable-factors-and-cluster-based covariance estimation (Factor+Cluster) with several competing methods that typically decompose the covariance matrix estimation into the sum of a low-rank common covariance matrix, formed by observable factors or latent factors that can be obtained through PCA, and a regularized remaining covariance matrix. These competitors include: the latent factor model with a cluster-based remaining covariance matrix (PCA+Cluster) as detailed in Remark~\ref{rmk.pcacluster}, the observable factor model with a sparse remaining (i.e., idiosyncratic) covariance matrix in \cite{fan2011high} (Factor+Sparsity), 
the latent factor model with a sparse idiosyncratic covariance matrix, referred to as the POET estimator in \cite{fan2013large}, the observable factor model with an idiosyncratic covariance matrix estimated by the linear shrinkage method in \cite{ledoit2004well} (Factor+LS) or the nonlinear shrinkage method in \cite{ledoit2012nonlinear} (Factor+NLS). For comparison, we also employ the sample covariance matrix estimator (Sample) and the covariance estimator based on the latent two-layer factor model with common strong factors and cluster-specific weak factors in \cite{zhang2023factor} (PCA+Weak). It is worth noting that, while the error covariance matrix $\bSigma_e$ is not required to have any specific structural assumption in \cite{zhang2023factor}, we assume that $\bSigma_e$ is diagonal, which significantly improves the performance.

\begin{table}[htbp]
  \centering
  \caption{Annualized performance measures (in percent) for eight covariance matrix estimators. The best performance is highlighted in bold.}
    \begin{tabular}{ccccccccc}
    \toprule
          &       & \multicolumn{3}{c}{Unconstrained} &       & \multicolumn{3}{c}{Constrained} \\
\cmidrule{3-5}\cmidrule{7-9}          &       & AV    & SD    & IR    &       & AV    & SD    & IR \\
    \midrule
    Factor+Cluster &       & \textbf{13.81 } & \textbf{9.17 } & \textbf{1.51 } &       & \textbf{10.75 } & \textbf{8.57 } & \textbf{1.25 } \\
    PCA+Cluster &       & 10.69  & 10.56  & 1.01  &       & 9.60  & 9.83  & 0.98  \\
    Factor+Sparsity &       & 0.17  & 10.99  & 0.01  &       & 8.62  & 8.62  & 0.99  \\
    POET  &       & 6.65  & 16.56  & 0.40  &       & 9.91  & 8.64  & 1.15  \\
    Factor+LS &       & 10.55  & 9.28  & 1.14  &       & 8.35  & 8.62  & 0.97  \\
    Factor+NLS &       & 10.61  & 9.26  & 1.15  &       & 8.36  & 8.59  & 0.98  \\
    PCA+Weak &       & 8.02  & 11.56 & 0.69  &       & 7.62  & 8.84  & 0.86 \\
    Sample &       & 13.14  & 29.47  & 0.45  &       & 7.56  & 8.71  & 0.87  \\
    \bottomrule
    \end{tabular}%
  \label{Table2}%
\end{table}%

\begin{figure}[htbp]
	\centering
	\includegraphics[height=18cm, width=15cm]{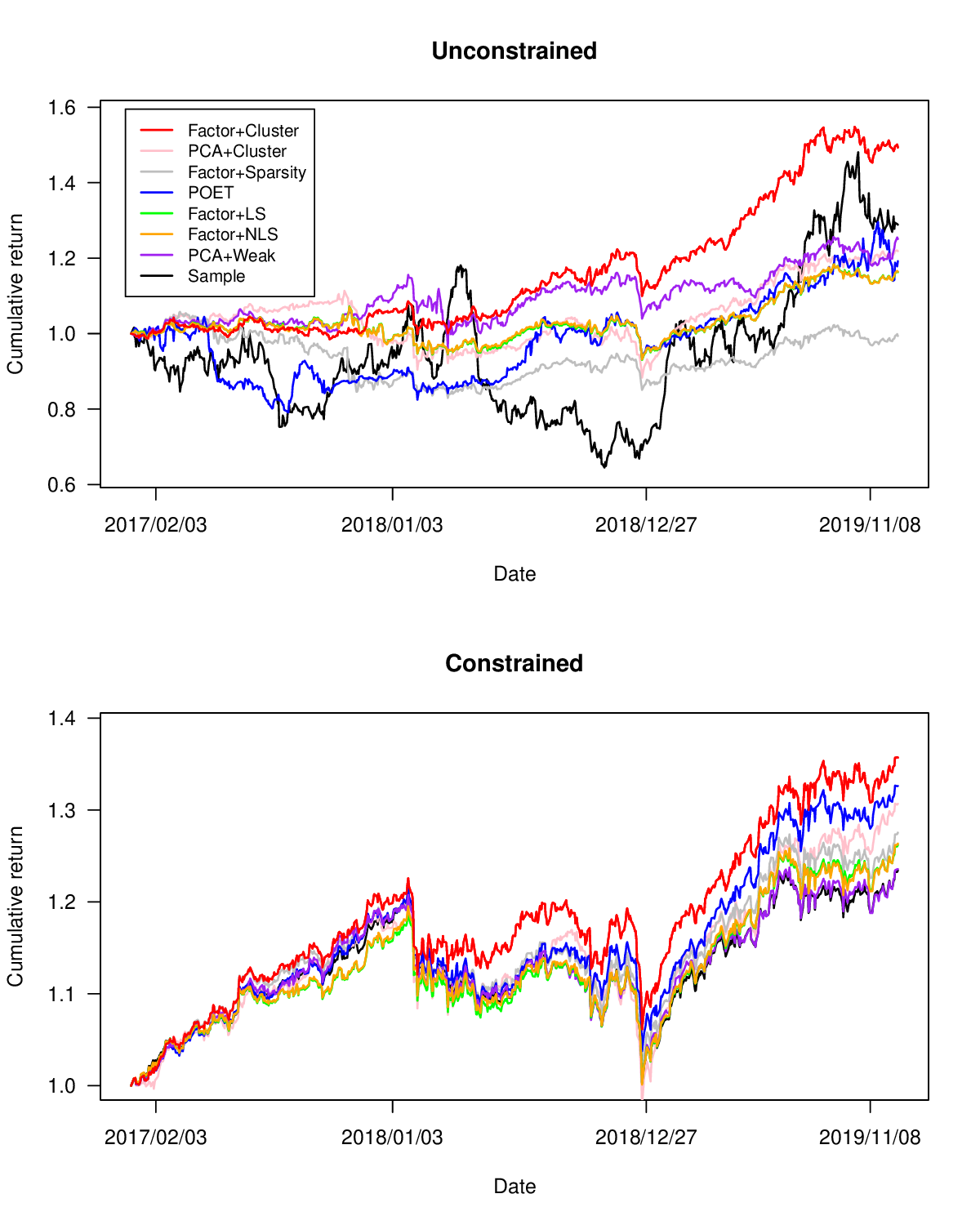}
	\caption{Out-of-sample cumulative returns of constrained and unconstrained minimal variance portfolios.}
	\label{Cumulative}
\end{figure}

Table~\ref{Table2} reports the annualized AV, SD, and IR of the unconstrained and constrained minimum variance portfolios, constructed using eight covariance matrix estimators for comparison. Several conclusions can be drawn from this table.  
%the observable factor model with cluster-structured idiosyncratic covariance matrix and its competitors. From this table, we have several observations.
First, the constructed minimum variance portfolios by our method perform the best in terms of AV, SD, and IR under both constrained and unconstrained cases. Notably, the IR for the unconstrained minimum variance portfolios of our proposed method is 31.3\% higher than that of the Factor+NLS method, which achieves the second-best performance.
The substantial superiority of our estimator is likely due to its latent cluster structure, which effectively alleviates the curse of dimensionality while preserving important cross-sectional information after removing common factors.
Second, the significant outperformance of our proposal over the PCA+Cluster method highlights the advantages of incorporating observable factors in the modelling context. See also Remark~\ref{rmk.pcacluster}.
Third, the performance of the sample estimator severely deteriorates in terms of SD under the unconstrained case. This is expected due to the error accumulation in the high-dimensional case, underscoring the necessity of choosing a proper structure for large covariance matrix estimation.
Finally, we plot the out-of-sample cumulative returns of the unconstrained and constrained minimum variance portfolios constructed using all the competing methods in Figure~\ref{Cumulative}. It is apparent that our constructed minimal variance portfolios achieve the highest cumulative returns under both the constrained and unconstrained cases during the test period. This finding is consistent with the superior AV performance of our estimator in Table \ref{Table2}.

\section{Discussion}
\label{sec:discussion}
We identify several important directions for future research.
Our paper assumes the large covariance matrix to be static with constant entries. However, in practical applications, time series variables often exhibit smooth structural changes over a long time period. Therefore, it is of interest to estimate the large dynamic covariance matrix by allowing the entries to vary smoothly with a certain variable or time, as done in \cite{chen2016dynamic} and \cite{wang2021nonparametric}. One possible extension is to implement a kernel-weighted estimation approach to estimate time-varying factor loadings and identify the dynamic latent cluster structure, based on which we can obtain dynamic covariance matrix estimation.
Another interesting extension is to consider the robust estimation of large covariance matrix, as it is well known that financial time series commonly suffer from heavy tails. Motivated by \cite{fan2019robust}, one can potentially employ adaptive Huber loss minimization or Kendall's tau correlation to estimate the joint covariance matrix of the observed data and the factors, then use it to estimate the residual covariance matrix and based on which develop a robust procedure to discover the latent cluster structure, and finally recover the covariance matrix of the observed data. These topics are beyond the scope of the current paper and will be pursued elsewhere.

\begin{appendix}

\section*{Appendix}

This appendix contains the empirical evidence against the sparsity assumption in Section~\ref{ap.sparse} and the technical proofs of the theoretical results in Section~\ref{ap.proof}.

\appendix

\setcounter{equation}{0}
\renewcommand{\theequation}{A.\arabic{equation}}
\section{Empirical evidence against the sparsity assumption}
\label{ap.sparse}
In \cite{fan2013large}, the %idiosyncratic 
covariance matrix $\bSigma_u=(\Sigma_{u,ij})_{p \times p}$ is assumed to be approximately sparse, i.e., the measure of the sparsity level,
\begin{equation}\label{m_p}
    m_p = \max_{i \in [p]} \sum_{j\in [p]} |\Sigma_{u,ij}|^{\kappa}~~\text{for some}~~ \kappa \in [0,1),
\end{equation}
does not grow too fast as $p \rightarrow \infty$. This encompasses the exact sparsity assumption ($\kappa=0$), under which $m_p = \max_{i\in [p]} \sum_{j\in [p]} I(\Sigma_{u,ij}\not = 0)$ corresponds to the maximum number of non-zero entries in each row. The exact sparsity assumption for $\bSigma_u$ is also considered in \cite{fan2011high}.

\begin{figure}[!htbp]
    \centering
    \begin{subfigure}
        \centering
        \includegraphics[width=8.5cm]{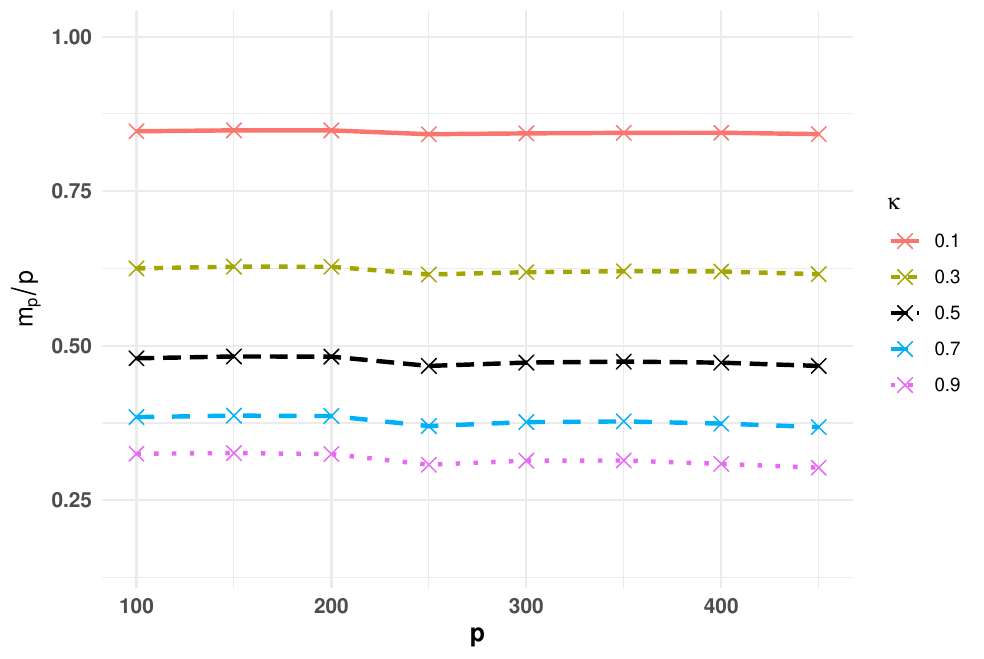}
        %\caption{Fama--French 5 common factors}
        %\label{fig:FF-5}
    \end{subfigure}
    %\hfill
    \begin{subfigure}
        \centering
        \includegraphics[width=8.5cm]{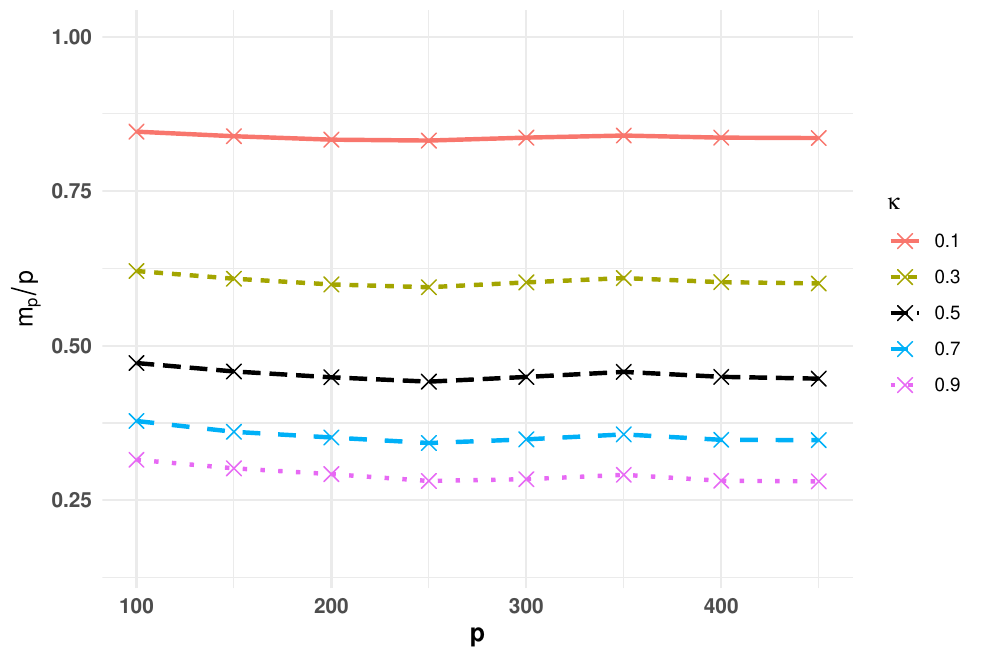}
        %\caption{Common factors estimated by PCA}
        %\label{fig:PCA}
    \end{subfigure}
    \caption{Plots of $\hat m_p/p$ as functions of $p=100, 150, 200, 250, 300, 350, 400, 450$ over different values of $\kappa=0.1, 0.3, 0.5, 0.7, 0.9$ for $\widecheck{\bSigma}_u$ after removing the common components using the Fama--French five factors (left) and the three latent factors estimated by PCA (right).}
    \label{fig:m_p/p}
\end{figure}

However, such a sparsity assumption may not hold in practice. Based on the S$\&$P 500 stock return dataset analyzed in Section \ref{sec.real} and \cite{zhang2023factor}, we conduct empirical analysis to check the validity of the sparsity assumption for $\bSigma_u$. Specifically, we estimate the factor loadings using both the Fama--French five factors and the three latent factors estimated by PCA \cite[]{fan2013large}. After removing the common components,  we obtain the sample covariance matrix of residuals $\widehat{\mathbf{u}}_t$ as $\widecheck{\bSigma}_u=(\widecheck \Sigma_{u,ij})_{p \times p}$ and estimate $m_p$ as $\hat m_p$ by replacing $\bSigma_u$ in \eqref{m_p} with $\widecheck\bSigma_u$ for different values of the parameter $\kappa \in [0,1)$. Figure \ref{fig:m_p/p} plots the ratio $\hat m_p / p$ as a function of $p$ ranging from 100 to 450, which is randomly selected from a pool of 477 stocks, %100, 150, 200, 250, 300, 350, 400$ and $450$ 
using both observable and latent factor models. It is obvious that, as $p$ increases,  $\hat m_p/p$ remains relatively steady regardless of different values of $\kappa$, providing empirical evidence against the approximate sparsity assumption with $m_p = o(p)$.

Additionally, we formulate the validation of the sparsity assumption as a problem of simultaneously testing $p(p-1)/2$ hypothesis for the off-diagonal entries of $\bSigma_u.$
We apply the multiple testing procedure in \cite{cai2016large} to $\widecheck{\bSigma}_u$ at a target false discovery rate level of $0.2$. Among $p(p-1)/2$ off-diagonal entries in ${\bSigma}_u$ with $p=477,$ $42.01\%$ and $40.23\%$ are identified as nonzero ones after removing the common components using the Fama--French five factors and the three latent factors estimated by PCA, respectively. This result provides further empirical evidence against the sparsity assumption for $\bSigma_u$ using the financial dataset.

\setcounter{equation}{0}
\renewcommand{\theequation}{B.\arabic{equation}}
\section{Proofs of theoretical results in Section~\ref{sec.theory}}
\label{ap.proof}

Throughout, we use $C$ to denote a generic positive constant that may be different in different places.

\subsection{Proof of Theorem \ref{thm1}}
To prove Theorem~\ref{thm1}, we first present some technical lemmas along with their proofs.

\begin{lemma}\label{Lemma0}(Lemma~2 in \cite{fan2013large}) 
Let $\bA$ and $\mathbf{B}$ be semi-positive definite matrices. If there exists a sequence $c_N>0$ such that $\lambda_{\min}(\mathbf{B}) > c_N$ and $\|\bA - \mathbf{B} \| =o_{\p}(c_N),$ then $\lambda_{\min}(\bA) > c_N/2,$ and $\|\bA^{-1} - \mathbf{B}^{-1} \| = O_{\p}(c_N^{-2})\|\bA - \mathbf{B} \|.$

\end{lemma}

\begin{lemma}\label{Lemma_Nagaev}
If Conditions~\ref{cond1}--\ref{cond2} hold, then there exists a constant $C > 0$ (only depending on $l$ and $\tau$ in Condition~\ref{cond2}) such that, for any $x \in (0,1),$
\begin{equation*}
    \eP\Big( \max_{i,j \in [p]}\Big|\frac{1}{T}\sum_{t=1}^{T}u_{it}u_{jt} - \E(u_{it}u_{jt}) \Big| \ge x \Big) \le C p^2 \big\{   T (Tx)^{-l} + \exp(-C T x^2)\big\}.
\end{equation*}
\end{lemma}

%\subsection{Proof of Lemma~\ref{Lemma_Nagaev}}
\noindent  {\sc Proof.}
Condition~\ref{cond2}(i) implies that $\max_{t \in [T]}\max_{i,j \in [p]} \eP(|u_{it}u_{jt} - \E(u_{it}u_{jt})| > x ) = O(x^{-l-s})$ as $x \rightarrow \infty.$ 
By the Fuk--Nagaev inequality in Theorem 6.2 of \cite{rio2017asymptotic}, for any $x > 0$, we have
\begin{equation*}
    \eP\Big( \Big|\frac{1}{T}\sum_{t=1}^{T}u_{it}u_{jt} - \E(u_{it}u_{jt}) \Big| \ge x \Big) \le C \big\{   T (Tx)^{-l} + \exp(-C T x^2)\big\}.
\end{equation*}
Applying the union bound of probability, we obtain that 
\begin{equation*}
     \eP\Big( \max_{i,j \in [p]}\Big|\frac{1}{T}\sum_{t=1}^{T}u_{it}u_{jt} - \E(u_{it}u_{jt}) \Big| \ge x \Big) 
     \le p^2\,  \eP\Big( \Big|\frac{1}{T}\sum_{t=1}^{T}u_{it}u_{jt} - \E(u_{it}u_{jt}) \Big| \ge x \Big),
\end{equation*}
which completes the proof.
\qed

\begin{lemma}\label{Lemma_maxnorm}
Let $\omega(p,T) = \max\{p^{2/l} T^{-(l-1)/l}, (T^{-1} \log p)^{1/2} \}$ and $\omega(K,T) = \max\{K^{2/l} T^{-(l-1)/l}, (T^{-1} \log K)^{1/2} \}$. If Conditions~\ref{cond1}--\ref{cond3} hold, then it holds that:
\begin{enumerate}
    \item[$\mathrm{(i)}$] $\max_{i,j \in [p]} \big|T^{-1}\sum_{t=1}^{T}u_{it}u_{jt} - \E\big(u_{it}u_{jt}\big)\big| = O_{\p}\big(\omega(p,T)\big),$
    \item[$\mathrm{(ii)}$] $\max_{i,j \in [p]} \big|T^{-1}\sum_{t=1}^{T}e_{it}e_{jt} - \E\big(e_{it}e_{jt}\big)\big| = O_{\p}\big(\omega(p,T)\big),$
    \item[$\mathrm{(iii)}$] $\max_{i,j \in [K]} \big|T^{-1}\sum_{t=1}^{T}z_{it}z_{jt} - \E\big(z_{it}z_{jt}\big)\big| = O_{\p}\big(\omega(K,T)\big),$
    \item[$\mathrm{(iv)}$] $\max_{i,j \in [r]} \big|T^{-1}\sum_{t=1}^{T}f_{it}f_{jt} - \E\big(f_{it}f_{jt}\big)\big| = O_{\p}\big(1/\sqrt{T}\big),$
    \item[$\mathrm{(v)}$] $\max_{i\in [K], j \in [p]} \big|T^{-1}\sum_{t=1}^{T}z_{it}e_{jt} \big| = O_{\p}\big(\omega(p,T)\big),$
    \item[$\mathrm{(vi)}$] $\max_{i\in [r], j \in [p]} \big|T^{-1}\sum_{t=1}^{T}f_{it}u_{jt} \big| = O_{\p}\big(\omega(p,T)\big),$
    \item[$\mathrm{(vii)}$] $\max_{i\in [r], j \in [p]} \big|T^{-1}\sum_{t=1}^{T}f_{it}e_{jt} \big| = O_{\p}\big(\omega(p,T)\big),$
    \item[$\mathrm{(viii)}$] $\max_{i\in [r], j \in [K]} \big|T^{-1}\sum_{t=1}^{T}f_{it}z_{jt} \big| = O_{\p}\big(\omega(K,T)\big).$
\end{enumerate}
\end{lemma}

\noindent  {\sc Proof.}
The proof is similar to that of Lemma~\ref{Lemma_Nagaev} by replacing $u_{it}u_{it}$ with related terms and some  calculations, and thus is omitted.
\qed

\begin{lemma}\label{Lemma_bhat}
If Conditions~\ref{cond1}--\ref{cond3} hold, then it holds that:
\begin{enumerate}
    \item[$\mathrm{(i)}$] $\max_{i \in [p]}\|\widehat{\mathbf{b}}_i-\mathbf{b}_i\| = O_{\p}( \omega(p,T) ),$
    \item[$\mathrm{(ii)}$] $\max_{i \in [p]} T^{-1} \sum_{t=1}^T|\widehat{u}_{it} - u_{it} |^2 = O_{\p}(\omega(p,T)^2).$
\end{enumerate}
\end{lemma}

%\subsection{Proof of Lemma \ref{Lemma_bhat}}
\noindent  {\sc Proof.}
(i) Note that  
$\widehat{\mathbf{b}}_i - \mathbf{b}_i = (T^{-1}\sum_{t=1}^{T}\mathbf{f}_t\mathbf{f}_t^{\T})^{-1} (T^{-1}\sum_{t=1}^{T}\mathbf{f}_t u_{it})$, then $$\|\widehat{\mathbf{b}}_i - \mathbf{b}_i\|^2 = \Big(\frac{1}{T}\sum_{t=1}^{T}u_{it}\mathbf{f}_t^{\T} \Big) \Big(\frac{1}{T}\sum_{t=1}^{T}\mathbf{f}_t\mathbf{f}_t^{\T}\Big)^{-2}\Big(\frac{1}{T}\sum_{t=1}^{T}\mathbf{f}_t u_{it}\Big).$$ 
By Lemma \ref{Lemma0}, $\eP(\lambda_{\min}(T^{-1}\sum_{t=1}^{T}\mathbf{f}_t\mathbf{f}_t^{\T}) \ge \lambda_{\min}(\E(\mathbf{f}_t\mathbf{f}_{t}^{\T}))/2) \ge \eP(\|T^{-1}\sum_{t=1}^{T}\mathbf{f}_t\mathbf{f}_t^{\T} - \E(\mathbf{f}_t \mathbf{f}_t^{\T}) \| \le \lambda_{\min}( \E(\mathbf{f}_t \mathbf{f}_t^{\T}))/2)$.  By Lemma~\ref{Lemma_maxnorm}(iv) and (vi), for any $\epsilon > 0$, there exists a positive constant $C_1$ such that each of the events $A_1 = \{\lambda_{\min}(T^{-1}\sum_{t=1}^{T}\mathbf{f}_t\mathbf{f}_t^{\T}) \ge \lambda_{\min}(\E(\mathbf{f}_t\mathbf{f}_{t}^{\T}))/2 \}$ and $A_2 = \{ \max_{l\in [r], i \in [p]}\big| T^{-1}\sum_{t=1}^{T}f_{lt}u_{it}\big| \le C_1 \omega(p,T) \}$  occurs with probability at least $1-\epsilon$ when $T$ is large enough.
Thus, under the event $A_1 \cap A_2$, we have 
\begin{equation*}
    \|\widehat{\mathbf{b}}_i - \mathbf{b}_i \|^2 \le \frac{4}{\lambda_{\min}^2( \E(\mathbf{f}_t \mathbf{f}_t^{\T}) )} \sum_{l=1}^{r}\Big(\frac{1}{T}\sum_{t=1}^Tf_{lt}u_{it}\Big)^2 \le \frac{4r}{\lambda_{\min}^2( \E(\mathbf{f}_t \mathbf{f}_t^{\T}) )}  \max_{l\in [r], i\in [p]}\Big| \frac{1}{T}\sum_{t=1}^{T}f_{lt}u_{it}\Big|^2 \le C_2^2 \omega(p,T)^2
\end{equation*}
with $C_2 = 2 C_1\sqrt{r} /\lambda_{\min}(\E(\mathbf{f}_t \mathbf{f}_t^{\T})) $. Then, with probability at least $1-2\epsilon$, we have $\|\widehat{\mathbf{b}}_i - \mathbf{b}_i\| \le C_2 \omega(p,T)$. Since $\epsilon$ is arbitrary, we have $\|\widehat{\mathbf{b}}_i - \mathbf{b}_i\| = O_{\p}(\omega(p,T))$, which completes the proof of (i). 

(ii) By Lemma~\ref{Lemma_maxnorm}(iv), clearly, $T^{-1}\sum_{t=1}^{T}\|\mathbf{f}_t \|^{2} \le \E\|\mathbf{f}_t\|^2 + r \max_{i \in [r]} \big|T^{-1}\sum_{t=1}^{T}f_{it}^2 - \E\big(f_{it}^2\big)\big| = O_{\p}(1)$. Then, it follows that
\begin{flalign*}
  &&  \max_{i \in [p]} \frac{1}{T} \sum_{t=1}^{T}| \widehat{u}_{it} - u_{it}|^2 \le \max_{i \in [p]} \frac{1}{T} \sum_{t=1}^{T} \|\mathbf{f}_t \|^2 \|\widehat{\mathbf{b}}_i -\mathbf{b}_i \|^2 = O_{\p}(\omega(p,T)^2). &&\qed
\end{flalign*}

\begin{lemma}\label{Lemma_concentration_cov}
If Conditions \ref{cond1}--\ref{cond3} hold, then it holds that
$$\max_{i,j,l \in [p]} |\widehat{\operatorname{cov}}(\widehat{u}_{it} - \widehat{u}_{jt}, \widehat{u}_{lt}) - \operatorname{cov}(u_{it} - u_{jt}, u_{lt})| = O_{\p} (\omega(p,T)).
$$
\end{lemma}

%\subsection{Proof of Lemma \ref{Lemma_concentration_cov}}
\noindent {\sc Proof.} 
Note that 
\begin{align*}
   & \max_{i,j,l \in [p]} |\widehat{\text{cov}}(\widehat{u}_{it} - \widehat{u}_{jt}, \widehat{u}_{lt}) - \text{cov}(u_{it} - u_{jt}, u_{lt})| \\
     &\le \max_{i,j,l \in [p]} |\widehat{\text{cov}}(\widehat{u}_{it} - \widehat{u}_{jt}, \widehat{u}_{lt}) - \widehat{\text{cor}}(u_{it} - u_{jt}, u_{lt})| + \max_{i,j,l \in [p]} |\widehat{\text{cov}}(u_{it} - u_{jt}, u_{lt}) - \text{cov}(u_{it} - u_{jt}, u_{lt})|\\
     &:= I_1 + I_2.
\end{align*}
For the term $I_1$, we have
\begin{flalign*}
\qquad &\max_{i,j,l \in [p]} |\widehat{\text{cov}}(\widehat{u}_{it} - \widehat{u}_{jt}, \widehat{u}_{lt}) - \widehat{\text{cov}}(u_{it} - u_{jt}, u_{lt})| \\
&= \max_{i,j,l \in [p]} \Big| \frac{1}{T}\sum_{t=1}^{T}(\widehat{u}_{it}\widehat{u}_{lt} - \widehat{u}_{jt}\widehat{u}_{lt}) - \frac{1}{T}\sum_{t=1}^{T}({u}_{it}{u}_{lt} - {u}_{jt}{u}_{lt}) \Big| \\
&\le \max_{i,l \in [p]}\Big| \frac{1}{T}\sum_{t=1}^{T}(\widehat{u}_{it}\widehat{u}_{lt} - {u}_{it}{u}_{lt}) \Big| + \max_{j,l \in [p]}\Big| \frac{1}{T}\sum_{t=1}^{T}(\widehat{u}_{jt}\widehat{u}_{lt} - {u}_{jt}{u}_{lt}) \Big|. &&
\end{flalign*}
By Lemma~\ref{Lemma_maxnorm}(i) and Lemma~\ref{Lemma_bhat}(ii), for any $\epsilon > 0$, there exists a positive constant $C_3$ such that, when $T$ is large enough, each of the events $A_3 = \{ \max_{i \in [p]}|T^{-1}\sum_{t=1}^{T}u_{it}^2 - \E(u_{it}^2)|\le \max_{i \in [p]}\E(u_{it}^2)/9\}$ and $A_4 = \{ \max_{i \in [p]} T^{-1}\sum_{t=1}^{T}|\widehat{u}_{it} - u_{it}|^2 \le C_3 \omega(p,T)^2   \}$  occurs with probability at least $1-\epsilon$.
Thus, under the event $A_3 \cap A_4$, by the Cauchy--Schwarz inequality, we have 
\begin{align*}
    \max_{i,l \in [p]}\Big| \frac{1}{T}\sum_{t=1}^{T}(\widehat{u}_{it}\widehat{u}_{lt} - {u}_{it}{u}_{lt}) \Big| 
    &\le \max_{i,l \in [p]} \Big| \frac{1}{T}\sum_{t=1}^{T}(\widehat{u}_{it} - {u}_{it}) (\widehat{u}_{lt} - {u}_{lt}) \Big| + 2 \max_{i,l \in [p]}  \Big| \frac{1}{T}\sum_{t=1}^{T}u_{it} (\widehat{u}_{lt} - {u}_{lt}) \Big| \\
    &\le \max_{i \in [p]}\frac{1}{T}\sum_{t=1}^{T}(\widehat{u}_{it} - u_{it})^2 + 2 \sqrt{\max_{i \in [p]}\frac{1}{T}\sum_{t=1}^{T}u_{it}^2}\, \sqrt{\max_{i \in [p]}\frac{1}{T}\sum_{t=1}^{T}(\widehat{u}_{it} - u_{it})^2}\\
    &\le C_3\omega(p,T)^2 + 2\sqrt{10/9}\sqrt{\max_{i \in [p]}\E(u_{it}^2)}\sqrt{C_3 \omega(p,T)^2}.
\end{align*}
Since $\omega(p,T) = o(1)$ and $\max_{i \in [p]}\E(u_{it}^2) = O(1)$ by Condition \ref{cond2}, choosing $C_4 > 4\sqrt{C_3 \max_{i\in [p]}\E(u_{it}^2)}$, we then have $C_4 \omega(p,T) \ge C_3\omega(p,T)^2 + 2\sqrt{10/9}\sqrt{\max_{i \in [p]}\E(u_{it}^2)}\sqrt{C_3 \omega(p,T)^2}$ for large $T$. Thus, with probability at least $1-2\epsilon$,  $\max_{i,l \in [p]}\big| T^{-1}\sum_{t=1}^{T}(\widehat{u}_{it}\widehat{u}_{lt} - {u}_{it}{u}_{lt}) \big|  \le C_4 \omega(p,T)$. Since $\epsilon$ is arbitrary, we have $\max_{i,l \in [p]}\big|T^{-1} \sum_{t=1}^{T}(\widehat{u}_{it}\widehat{u}_{lt} - {u}_{it}{u}_{lt}) \big| = O_{\p}(\omega(p,T))$. Similarly, we can obtain that $\max_{j,l \in [p]}\big| T^{-1}\sum_{t=1}^{T}(\widehat{u}_{jt}\widehat{u}_{lt} - {u}_{jt}{u}_{lt}) \big| = O_{\p}(\omega(p,T))$. Thus, $\max_{i,j,l \in [p]} |\widehat{\text{cov}}(\widehat{u}_{it} - \widehat{u}_{jt}, \widehat{u}_{lt}) - \widehat{\text{cov}}(u_{it} - u_{jt}, u_{lt})| = O_{\p}(\omega(p,T))$.

For the term $I_2$, a simple calculation and Lemma~\ref{Lemma_maxnorm}(i) give
\begin{flalign*}
\qquad
&\max_{i,j,l \in [p]} |\widehat{\text{cov}}(u_{it} - u_{jt}, u_{lt}) - \text{cov}(u_{it} - u_{jt}, u_{lt})| \\
&= \max_{i,j,l \in [p]} \Big|\frac{1}{T}\sum_{t=1}^{T}(u_{it}u_{lt} - u_{jt}u_{lt}) - \E(u_{it}u_{lt} - u_{jt}u_{lt}) \Big| \\
&\le \max_{i,l \in [p]}\Big|\frac{1}{T}\sum_{t=1}^{T}u_{it}u_{lt} - \E(u_{it}u_{lt}) \Big| + \max_{j,l \in [p]}\Big|\frac{1}{T}\sum_{t=1}^{T}u_{jt}u_{lt} - \E(u_{jt}u_{lt}) \Big| =  O_{\p}(\omega(p,T)). &&
\end{flalign*}
Thus, $\max_{i,j,l \in [p]} |\widehat{\text{cov}}(\widehat{u}_{it} - \widehat{u}_{jt}, \widehat{u}_{lt}) - \text{cov}(u_{it} - u_{jt}, u_{lt})| = O_{\p} (\omega(p,T))$. The proof is complete. 
\qed

\begin{lemma}\label{Lemma_concentration_cor}
If Conditions~\ref{cond1}--\ref{cond3} hold, then it holds that
$$
\max_{i,j,l \in [p]} |\widehat{\operatorname{cor}}(\widehat{u}_{it} - \widehat{u}_{jt}, \widehat{u}_{lt}) - \operatorname{cor}(u_{it} - u_{jt}, u_{lt})| = O_{\p} (\omega(p,T)).
$$
\end{lemma}

%\subsection{Proof of Lemma \ref{Lemma_concentration_cor}}
\noindent {\sc Proof.} 
Let $d = p(p+1)/2$, $\mathbf{v} = (v_1, \dots, v_d)^{\T}$ be a random vector consisting of  $u_{it} - u_{jt}$ for $1\le i <j \le p$ and $u_{lt}$ for $l \in [p]$. Similarly, let $\widehat{\mathbf{v}} = (\widehat{v}_1, \dots, \widehat{v}_d)^{\T}$ by replacing $u_{it}$ in $\mathbf{v}$ by $\widehat{u}_{it}$.  Denote $s_{ij} = \operatorname{cov}(v_i, v_j)$ and $r_{ij} = \operatorname{cor}(v_i, v_j)$, and similarly  $\widehat{s}_{ij} = \widehat{\operatorname{cov}}(\widehat{v}_i, \widehat{v}_j)$ and $\widehat{r}_{ij} = \widehat{\operatorname{cor}}(\widehat{v}_i, \widehat{v}_j)$. It suffices to prove that
\begin{equation}\label{Eq_concentration_cor}
\max_{i,j \in [d]}|\widehat{\operatorname{cor}}(\widehat{v}_i, \widehat{v}_j) - \operatorname{cor}(v_i, v_j) | \le 2\max_{i,j \in [d]} \frac{|\widehat{\operatorname{cov}}(\widehat{v}_i, \widehat{v}_j) - \operatorname{cov}(v_i, v_j)|}{\sqrt{\operatorname{var}(v_i)\operatorname{var}(v_j)}}.
\end{equation}
Note that $\widehat{u}_{it} = u_{it} - (\sum_{t=1}^{T}u_{it}\mathbf{f}_t^{\T})(\sum_{t=1}^{T}\mathbf{f}_t\mathbf{f}_t^{\T})^{-1} \mathbf{f}_t$ is linear in $u_{it}$ for $i \in [p]$, so is $\widehat{v}_{i}$ in $v_i$ for $i \in [d]$, which implies that both sides of \eqref{Eq_concentration_cor}  are scale-invariant in $v_i$. Without loss of generality, we assume that $\operatorname{var}(v_i) = 1$. Then, $\operatorname{cov}(v_i, v_j) = \operatorname{cor}(v_i, v_j)$ for $i,j \in [d]$ and
\begin{align*}
    |\widehat{r}_{ij} - r_{ij}| 
    &=  \big|\widehat{r}_{ij}\big(1 - \sqrt{\widehat{s}_{ii}\widehat{s}_{jj}}\big) + \widehat{s}_{ij} - s_{ij}\big|
    \le |\widehat{r}_{ij}| \big|1 - \sqrt{\widehat{s}_{ii}\widehat{s}_{jj}}\big| + | \widehat{s}_{ij} - s_{ij}|\\
    &\le \max\{|1 - \widehat{s}_{ii}|, |1 - \widehat{s}_{jj}| \} + | \widehat{s}_{ij} - s_{ij}|\\
    &=  \max\{|s_{ii} - \widehat{s}_{ii}|, |s_{jj} - \widehat{s}_{jj}| \} + | \widehat{s}_{ij} - s_{ij}|\\
    &\le 2\max_{i,j \in [d]} | \widehat{s}_{ij} - s_{ij}|
\end{align*}
by noting $|\widehat{r}_{ij}| \le 1$ and $s_{ii} = s_{jj} = 1$. Thus $\max_{i,j \in [d]}|\widehat{r}_{ij} - r_{ij}| \le 2\max_{i,j \in [d]} | \widehat{s}_{ij} - s_{ij}|$ and \eqref{Eq_concentration_cor} holds. 

Clearly, it follows that
\begin{align*}
    \max_{i,j \in [d]}|\widehat{\operatorname{cov}}(\widehat{v}_i, \widehat{v}_j) - \operatorname{cov}(v_i, v_j)| 
    \le& \max_{i,j,l \in [p]} |\widehat{\operatorname{cov}}(\widehat{u}_{it} - \widehat{u}_{jt}, \widehat{u}_{lt}) - \operatorname{cov}(u_{it} - u_{jt}, u_{lt})|\\
    &+ \max_{i,j \in [p]}|\widehat{\operatorname{var}}(\widehat{u}_{it} - \widehat{u}_{jt}) - \operatorname{var}(u_{it} - u_{jt})| + \max_{l\in [p]}|\widehat{\operatorname{var}}(\widehat{u}_{lt}) - \operatorname{var}(u_{lt})|. 
\end{align*} 
Similar to the proof of Lemma \ref{Lemma_concentration_cov}, it is not hard to prove that $\max_{i,j \in [p]}|\widehat{\operatorname{var}}(\widehat{u}_{it} - \widehat{u}_{jt}) - \operatorname{var}(u_{it} - u_{jt})| = O_{\p}(\omega(p,T))$ and $\max_{l\in [p]}|\widehat{\operatorname{var}}(\widehat{u}_{lt}) - \operatorname{var}(u_{lt})| = O_{\p}(\omega(p,T))$. Thus $ \max_{i,j \in [d]}|\widehat{\operatorname{cov}}(\widehat{v}_i, \widehat{v}_j) - \operatorname{cov}(v_i, v_j)| = O_{\p}(\omega(p,T))$. Moreover, by Condition \ref{cond1}, there exists a constant $C>0$ such that $\operatorname{var}(u_{lt}) \ge \operatorname{var}(e_{lt})> C $ and $\operatorname{var}(u_{it} - u_{jt}) \ge \operatorname{var}(e_{it}) + \operatorname{var}(e_{jt})> C $ for all $i,j, l \in [p]$. Thus, by\eqref{Eq_concentration_cor}, we have
\begin{equation*}
    \max_{i,j,l \in [p]} |\widehat{\operatorname{cor}}(\widehat{u}_{it} - \widehat{u}_{jt}, \widehat{u}_{lt}) - \operatorname{cor}(u_{it} - u_{jt}, u_{lt})| \le  \max_{i,j \in [d]}|\widehat{\operatorname{cor}}(\widehat{v}_i, \widehat{v}_j) - \operatorname{cor}(v_i, v_j) | = O_{\p} (\omega(p,T)).
\end{equation*}
The proof is complete.
\qed

We are now ready to prove Theorem \ref{thm1}.

%\textbf{Proof of Theorem \ref{thm1}.}
Recall that $D_{ij} = \operatorname{sCOD}(i,j) = \max_{l \not= i,j}|\operatorname{cor}(u_{it} - u_{jt}, u_{lt})|$ and  $\widehat{D}_{ij} = \widehat{\operatorname{sCOD}}(i,j) = \max_{l \not= i,j}|\widehat{\operatorname{cor}}(\widehat{u}_{it} - \widehat{u}_{jt}, \widehat{u}_{lt})|$. Then, it follows from Lemma~\ref{Lemma_concentration_cor} that
\begin{align}
\begin{split}\label{Eq_deviation_Dij}
     \max_{i,j \in [p]} |\widehat{D}_{ij} - D_{ij} | 
     =&  \max_{i,j \in [p]} \Big| \max_{l\not= i,j}|\widehat{\text{cor}}(\widehat{u}_{it} - \widehat{u}_{jt}, \widehat{u}_{lt})| - \max_{l\not= i,j}|\text{cor}(u_{it} - u_{jt}, u_{lt})| \Big|\\
     \le& \max_{i,j,l \in [p]} |\widehat{\text{cor}}(\widehat{u}_{it} - \widehat{u}_{jt}, \widehat{u}_{lt}) - \text{cor}(u_{it} - u_{jt}, u_{lt})|
     = O_{\p}(\omega(p,T)),
\end{split}
\end{align}
where the inequality holds since $|\max_{i\in [p]}a_i - \max_{i\in [p]}b_i| \le \max_{i\in [p]}|a_i - b_i|$ for two nonnegative sequences  $\{a_i\}_{i=1}^{p}$ and $\{b_i\}_{i=1}^{p}$.

By Condition~\ref{cond6}, we can choose  $h_T$ satisfying $\omega(p,T) = o (h_T)$ and $ h_T = o(\delta_T)$. Let $A_5 = \big\{\max_{i,j \in [p]} | \widehat{D}_{ij} - D_{ij}| \le h_T \le \varsigma/2$\big\}. Then, under the event $A_5$, we have that $\varsigma/2 \le \widehat{D}_{(q)} \le \varsigma/2 + D_{(1)}$ for $m \in [q]$ and $0 \le \widehat{D}_{(m)} \le h_T$ for $m > q$. Since $\widehat{q} = \arg\max_{m \in [Q]} \big\{\widehat{D}_{(m)} + \delta_{T}\big\}/\big\{\widehat{D}_{(m+1)} + \delta_{T}\big\}$, then, under the event $A_5$, we have
\begin{equation}\label{Asy_true}
    \frac{\widehat{D}_{(\widehat{q})} + \delta_T}{\widehat{D}_{(\widehat{q}+1)} + \delta_T} \ge     \frac{\widehat{D}_{(q)} + \delta_T}{\widehat{D}_{(q+1)} + \delta_T} \ge \frac{\varsigma/2 + \delta_T}{h_T + \delta_T} \asymp \frac{\varsigma}{\delta_T}.
\end{equation}
On the one hand, under the event $A_5$, if $\widehat{q} < q$, we have
\begin{equation}\label{Asy_small}
    \frac{\widehat{D}_{(\widehat{q})} + \delta_T}{\widehat{D}_{(\widehat{q}+1)} + \delta_T} \le \frac{\varsigma/2 + D_{(1)} + \delta_T}{\varsigma/2 + \delta_T} \asymp \frac{D_{(1)}}{\varsigma}.
\end{equation}
Noting that $\delta_T D_{(1)} / \varsigma^2 \rightarrow 0$ by Condition \ref{cond6}, then $\eP(\widehat{q} < q|A_5) \rightarrow 0$ by  \eqref{Asy_true}--\eqref{Asy_small}. On the other hand, under the  event $A_5$, if $\widehat{q} > q$, it follows that
\begin{equation}\label{Asy_big}
    \frac{\widehat{D}_{(\widehat{q})} + \delta_T}{\widehat{D}_{(\widehat{q}+1)} + \delta_T} \le \frac{h_T + \delta_T}{\delta_T} \rightarrow 1.
\end{equation}
Then $\eP(\widehat{q} > q|A_5) \rightarrow 0$ by $\varsigma / \delta_T \rightarrow \infty$, \eqref{Asy_true}, and \eqref{Asy_big}. Thus, we can obtain $\eP(\widehat{q} = q | A_5) \rightarrow 1$. Notice that $\eP(A_5) \rightarrow 1$ by \eqref{Eq_deviation_Dij}. Therefore,  $\eP(\widehat{q} = q)\geq\eP(\widehat{q} = q | A_5)\,\eP(A_5) \rightarrow 1$. Recall that $E = \{(i,j) | D_{ij} < D_{(q)}, 1\le i < j \le p \}$ and $\widehat{E} = \{(i,j) | \widehat{D}_{ij} < \widehat{D}_{(\widehat{q})}, 1\le i < j \le p \}$.  Under the event $\{\widehat{q} = q\}$, clearly,  the estimated graph $(G, \widehat{E})$ is identical to $(G, E),$ which completes the proof.
\qed

\subsection{Proof of Theorem \ref{thm2}}
By Theorem \ref{thm1}, the latent cluster structure can be estimated consistently, i.e. $\widehat{\bA} = \bA \bP$, where $\bP$ is a $K \times K$ orthogonal permutation matrix. In what follows, we assume $\widehat{\bA} = \bA \bP.$
Note that 
\begin{flalign*}
\widehat{\bSigma}_u - \bSigma_u &= 
(\widetilde{\bSigma}_u - \bSigma_u) +(\widehat{\bSigma}_u - \widetilde{\bSigma}_u)\\
&= 
\widehat{\bA}(\widetilde{\bSigma}_z - \bP^{\T}\bSigma_z\bP)\widehat{\bA}^{\T} 
+ (\widetilde{\bSigma}_e - \bSigma_e) + 
\widehat{\bA}(\widehat{\bSigma}_z - \widetilde{\bSigma}_z)\widehat{\bA}^{\T}  + (\widehat{\bSigma}_e - \widetilde{\bSigma}_e),
\end{flalign*}
where $\widetilde{\bSigma}_u = \widehat{\bA}\widetilde{\bSigma}_z \widehat{\bA}^{\T} + \widetilde{\bSigma}_e$, $\widetilde{\bSigma}_z = T^{-1}\sum_{t=1}^{T}\widetilde{\bz}_t \widetilde{\bz}_t^{\T}$, $\widetilde{\bSigma}_e = \bI_{p} \circ (T^{-1}\sum_{t=1}^{T}\widetilde{\be}_t \widetilde{\be}_t^{\T})$, and $\circ$ stands for the Hadamard product of two matrices such that $\bI_{p}\circ \bSigma = \operatorname{diag}(\bSigma).$ Here
$\widetilde{\mathbf{z}}_t  = \bP^{\T}\mathbf{z}_t + (\widehat{\bA}^{\T}\widehat{\bA})^{-1}\widehat{\bA}^{\T}\mathbf{e}_t$ and $
\widetilde{\mathbf{e}}_t = (\mathbf{I}_{p} - \mathbf{H})\mathbf{e}_t$ are the estimators for $\bz_t$ and $\be_t$ when $\bu_t$ is observed directly,  with $\mathbf{H} = \widehat{\bA}(\widehat{\bA}^{\T}\widehat{\bA})^{-1}\widehat{\bA}^{\T}$ being a projection matrix satisfying $\mathbf{H}^{\T}\mathbf{H} = \mathbf{H}$ and $\|\mathbf{H} \| = 1$.

To prove Theorem~\ref{thm2}, we present some technical lemmas below. Specifically, Lemmas~\ref{Lemma_Sigma_Z_tilde}--\ref{Lemma_Sigma_E_tilde} quantify various norms of $\widetilde{\bSigma}_z - \bP^{\T}\bSigma_z\bP$ and $\widetilde{\bSigma}_e - \bSigma_e$, Lemmas~\ref{Lemma_CT}--\ref{Lemma_Sigma_E_hat}  establish bounds under different norms for $\widehat{\bSigma}_z - \widetilde{\bSigma}_z$ and $\widehat{\bSigma}_e - \widetilde{\bSigma}_e$, and
Lemmas~\ref{Lemma12}--\ref{Lemma13} help to bound $\|\widehat{\bSigma}_u^{-1} - \bSigma_u\|$.
The proofs of these lemmas are provided in the supplementary material.

\begin{lemma}\label{Lemma_Sigma_Z_tilde}
If Conditions \ref{cond1}--\ref{cond5} hold, then it holds that:
\begin{enumerate}
	\item[$\mathrm{(i)}$]   $\big\| \widetilde{\bSigma}_z - \bP^{\T}\bSigma_z\bP \big\| = O_{\p}\left( K\omega(p,T) + K/p \right),$
    \item[$\mathrm{(ii)}$]   $\big\| \widetilde{\bSigma}_z - \bP^{\T}\bSigma_z\bP \big\|_{\max} = O_{\p}\left(\omega(p,T) + K/p \right),$
    \item[$\mathrm{(iii)}$]   $\big\| \widetilde{\bSigma}_z - \bP^{\T}\bSigma_z\bP \big\|_{\operatorname{F}} = O_{\p}\big(K\omega(p,T) + K^{3/2}/p \big),$
	\item[$\mathrm{(iv)}$]  $\big\| \widetilde{\bSigma}_z^{-1} - (\bP^{\T}\bSigma_z\bP)^{-1} \big\| = O_{\p}\left( K\omega(p,T) + K/p \right).$
\end{enumerate}
\end{lemma}

\begin{lemma}\label{Lemma_Sigma_E_tilde}
If Conditions \ref{cond1}--\ref{cond5} hold, then it holds that:
\begin{enumerate}
  \item[$\mathrm{(i)}$]   $\big\| \widetilde{\bSigma}_e - \bSigma_e \big\| = O_{\p}\left( K\omega(p,T) + K/p \right),$
  \item[$\mathrm{(ii)}$]  $\big\| \widetilde{\bSigma}_e - \bSigma_e \big\|_{\max} = O_{\p}\left( K\omega(p,T) + K/p \right),$
  \item[$\mathrm{(iii)}$]  $\big\| \widetilde{\bSigma}_e^{-1} - \bSigma_e^{-1} \big\| = O_{\p}\left( K\omega(p,T) + K/p \right).$
\end{enumerate}
\end{lemma}

\begin{lemma}\label{Lemma_CT}
%Define $\bC_T = \widehat{\mathbf{B}} - \mathbf{B}$. 
If Conditions \ref{cond1}--\ref{cond5} hold, then it holds that:
\begin{enumerate}
    \item[$\mathrm{(i)}$] $\|\widehat{\mathbf{B}} - \mathbf{B} \|_{\operatorname{F}} = O_{\p}(\sqrt{p}\,\omega(p,T)),$
    \item[$\mathrm{(ii)}$] $\|\widehat{\mathbf{B}} - \mathbf{B} \|_{\max} = O_{\p}(\omega(p,T)).$
\end{enumerate}
\end{lemma}

\begin{lemma}\label{Lemma_Sigma_Z_hat}
If Conditions \ref{cond1}--\ref{cond5} hold, then it holds that:
\begin{enumerate}
   \item[$\mathrm{(i)}$]   $\big\| \widehat{\bSigma}_z - \widetilde{\bSigma}_z \big\| = o_{\p}\left( K\omega(p,T) + K/p \right),$
    \item[$\mathrm{(ii)}$]   $\big\| \widehat{\bSigma}_z - \widetilde{\bSigma}_z \big\|_{\max} = o_{\p}\left(\omega(p,T) + K/p \right),$
    \item[$\mathrm{(iii)}$]   $\big\| \widehat{\bSigma}_z - \widetilde{\bSigma}_z \big\|_{\operatorname{F}} = o_{\p}\big(K\omega(p,T) + K^{3/2}/p \big),$
	\item[$\mathrm{(iv)}$]  $\big\| \widehat{\bSigma}_z^{-1} - \widetilde{\bSigma}_z^{-1} \big\| = o_{\p}\left( K\omega(p,T) + K/p \right).$
\end{enumerate}
\end{lemma}

\begin{lemma}\label{Lemma_Sigma_E_hat}
If Conditions~\ref{cond1}--\ref{cond5} hold, then it holds that:
\begin{enumerate}
    \item[$\mathrm{(i)}$] $\|\widehat{\bSigma}_e - \widetilde{\bSigma}_e \| = o_{\p}(K\omega(p,T) + K/p),$
    \item[$\mathrm{(ii)}$] $\|\widehat{\bSigma}_e - \widetilde{\bSigma}_e \|_{\max} = o_{\p}(K\omega(p,T) + K/p),$
    \item[$\mathrm{(iii)}$] $\|\widehat{\bSigma}_e^{-1} - \widetilde{\bSigma}_e^{-1} \| = o_{\p}(K\omega(p,T) + K/p).$
\end{enumerate}
\end{lemma}

\begin{lemma}\label{Lemma12}
If Conditions \ref{cond1}--\ref{cond5} hold, then it holds that:
\begin{enumerate}[(i)]
	\item[$\mathrm{(i)}$] $\lambda_{\min}(\widehat{\bA}^{\T}\bSigma_e^{-1}\widehat{\bA}) \ge C\,p/K $ for some constant $C>0,$
	\item[$\mathrm{(ii)}$] $\big\| \big\{(\bP^{\T}\bSigma_z\bP)^{-1} + \widehat{\bA}^{\T}\bSigma_e\widehat{\bA} \big\}^{-1} \big\|  = O_{\p}(K/p)$.
\end{enumerate}
\end{lemma}

\begin{lemma}\label{Lemma13}
If Conditions \ref{cond1}--\ref{cond5} hold and $K\omega(p,T) = o(1)$, then it holds that:
\begin{enumerate}
    \item[$\mathrm{(i)}$] $\big\| \widehat{\bA}^{\T}\widehat{\bSigma}_e^{-1}\widehat{\bA} - \widehat{\bA}^{\T}\bSigma_e^{-1}\widehat{\bA}  \big\| = O_{\p}(p\,\omega(p,T) + 1),$
    \item[$\mathrm{(ii)}$] $\|\widehat{\mathbf{M}}\| = O_{\p}(K/p),$ where $\widehat{\mathbf{M}}= \big( \widehat{\bSigma}_z^{-1} + \widehat{\bA}^{\T}\widehat{\bSigma}_e^{-1}\widehat{\bA} \big)^{-1},$
    \item[$\mathrm{(iii)}$] $ \|\widehat{\bA} \widehat{\mathbf{M}} \widehat{\bA}^{\T} \widehat{\bSigma}_e^{-1} \| = O_{\p}(1).$
\end{enumerate}
\end{lemma}

%\noindent  \textbf{Proof of Theorem \ref{thm2}}\\
We are now ready to prove Theorem \ref{thm2}.

(i) Recall that $\widehat{\bSigma}_u = \widehat{\bA}^{\T}\widehat{\bSigma}_z \widehat{\bA} + \widehat{\bSigma}_e$. 
%By Lemma~\ref{Lemma_Sigma_Z_tilde}(iii) and Lemma~\ref{Lemma_Sigma_Z_hat}(iii), $\|\widehat{\bSigma}_z - \bP^{\T}\bSigma_z\bP \|_{\operatorname{F}} = O_{\p}(K\omega(p,T) + K^{3/2}/p)$.
A simple calculation leads to
\begin{align*}
\widehat{\bA}^{\T}\bSigma_u^{-1}\widehat{\bA} 
    &= \widehat{\bA}^{\T}\bSigma_e^{-1}\widehat{\bA} - \widehat{\bA}^{\T}\bSigma_e^{-1}\widehat{\bA}\big[ (\bP^{\T}\bSigma_z\bP)^{-1} + \widehat{\bA}^{\T}\bSigma_e^{-1}\widehat{\bA} \big]^{-1} \widehat{\bA}^{\T}\bSigma_e^{-1}\widehat{\bA}\\
    &= (\bP^{\T}\bSigma_z\bP)^{-1} - (\bP^{\T}\bSigma_z\bP)^{-1}\big[ (\bP^{\T}\bSigma_z\bP)^{-1} + \widehat{\bA}\bSigma_e^{-1}\widehat{\bA} \big]^{-1} (\bP^{\T}\bSigma_z\bP)^{-1},
\end{align*}
and it follows that
\begin{align}\label{Eq_Sigma_u_norm1}
\begin{split}
\|\widehat{\bA}^{\T}\bSigma_u^{-1}\widehat{\bA} \|
    &\le \| (\bP^{\T}\bSigma_z\bP)^{-1} \| + \big\|(\bP^{\T}\bSigma_z\bP)^{-1}\big\{ (\bP^{\T}\bSigma_z\bP)^{-1} + \widehat{\bA}\bSigma_e^{-1}\widehat{\bA} \big\}^{-1} (\bP^{\T}\bSigma_z\bP)^{-1}
    \big\|\\
    &\le  2\| (\bP^{\T}\bSigma_z\bP)^{-1} \| 
    = O_{\p}(1).
\end{split}
\end{align} 
Applying Lemma~\ref{Lemma_Sigma_Z_tilde}(iii), Lemma~\ref{Lemma_Sigma_Z_hat}(iii) and  \eqref{Eq_Sigma_u_norm1} yields that 
\begin{equation*}
    \|\widehat{\bA}(\widehat{\bSigma}_z - \bP^{\T}\bSigma_z\bP)\widehat{\bA}^{\T}\|^2_{\bSigma_u} 
    \le  p^{-1} \|\widehat{\bSigma}_z - \bP^{\T}\bSigma_z\bP \|^2_{\operatorname{F}} \| \widehat{\bA}^{\T}\bSigma_u^{-1}\widehat{\bA} \|^2 
    =O_{\p}(K^2\omega(p,T)^2/p + K^{3}/p^3),
\end{equation*}
which implies $\|\widehat{\bA}(\widehat{\bSigma}_z - \bP^{\T}\bSigma_z\bP)\widehat{\bA}^{\T}\|_{\bSigma_u} = O_{\p}(K\omega(p,T)/\sqrt{p} + (K/p)^{3/2})$. Moreover,  Lemma~\ref{Lemma_Sigma_E_tilde}(i) and Lemma~\ref{Lemma_Sigma_E_hat}(i) lead to $ \|\widehat{\bSigma}_e - \bSigma_e \| = O_{\p}(K\omega(p,T) + K/p)$, thus
\begin{equation*}
\|\widehat{\bSigma}_e - \bSigma_e \|_{\bSigma_u} = O_{\p}(p^{-1/2})\|\widehat{\bSigma}_e - \bSigma_e \|_{\operatorname{F}} =O_{\p}\big( \|\widehat{\bSigma}_e - \bSigma_e \|\big) = O_{\p}(K\omega(p,T) + K/p).
\end{equation*}
Combing the above results, we complete the proof of part (i) by
$\|\widehat{\bSigma}_u - \bSigma_u \|_{\bSigma_u} \le   \|\widehat{\bA}(\widehat{\bSigma}_z - \bP^{\T}\bSigma_z\bP)\widehat{\bA}^{\T}\|_{\bSigma_u}  + \|\widehat{\bSigma}_e - \bSigma_e \|_{\bSigma_u} = O_{\p}(K\omega(p,T) + K/p).$

(ii) It follows from Lemma~\ref{Lemma_Sigma_Z_hat}(ii), Lemma~\ref{Lemma_Sigma_E_hat}(ii), and  $\|\widehat{\bSigma}_u - \bSigma_u \|_{\max} 
     \le   \|\widehat{\bA}(\widehat{\bSigma}_z - \bP^{\T}\bSigma_z\bP)\widehat{\bA}^{\T}\|_{\max}  + \|\widehat{\bSigma}_e - \bSigma_e \|_{\max}$ that part (ii) holds.

(iii) By Lemma~\ref{Lemma_Sigma_E_tilde}(i) and Lemma~\ref{Lemma_Sigma_E_hat}(i), when $K\omega(p,T) + K/p = o(1)$, it follows that $\lambda_{\min}(\widehat{\bSigma}_e)$ is bounded away from $0$ with probability approaching $1$ as $\lambda_{\min}(\bSigma_e)$ is bounded away from $0$. Consequently, $\widehat{\bSigma}_u$ has a bounded inverse with probability approaching $1$. 
Recall that $\bSigma_u^{-1} = \bSigma_e^{-1} -  \bSigma_e^{-1}\widehat{\bA}\big\{ (\bP^{\T}\bSigma_z\bP)^{-1} + \widehat{\bA}^{\T}\bSigma_e^{-1}\widehat{\bA} \big\}^{-1} \widehat{\bA}^{\T}\bSigma_e^{-1}$, then
\begin{align*}%\label{Eq_Sigma_u_inv}
\begin{split}
\big\|  \widehat{\bSigma}_u^{-1} -  \bSigma_u^{-1}\big\| 
\le& \big\|   \widehat{\bSigma}_e^{-1} -   \bSigma_e^{-1}\big\| +
\big\|(\widehat{\bSigma}_e^{-1} - \bSigma_e^{-1})\widehat{\bA}\big( \widehat{\bSigma}_z^{-1} + \widehat{\bA}^{\T}\widehat{\bSigma}_e^{-1}\widehat{\bA} \big)^{-1} \widehat{\bA}^{\T}\widehat{\bSigma}_e^{-1}\big\|\\
&+ \big\|\bSigma_e^{-1}\widehat{\bA}\big( \widehat{\bSigma}_z^{-1} + \widehat{\bA}^{\T}\widehat{\bSigma}_e^{-1}\widehat{\bA} \big)^{-1} \widehat{\bA}^{\T}(\widehat{\bSigma}_e^{-1} - \bSigma_e^{-1})\big\| \\
&+ \big\|\bSigma_e^{-1}\widehat{\bA}\big[\big( \widehat{\bSigma}_z^{-1} + \widehat{\bA}^{\T}\widehat{\bSigma}_e^{-1}\widehat{\bA}\big)^{-1} - \big\{ (\bP^{\T}\bSigma_z\bP)^{-1} + \widehat{\bA}^{\T}\bSigma_e^{-1}\widehat{\bA}\big\}^{-1} \big] \widehat{\bA}^{\T}\bSigma_e^{-1}\big\|\\
:=& \|\mathbf{L}_1\| + \|\mathbf{L}_{2}\| + \|\mathbf{L}_{3}\| + \|\mathbf{L}_{4}\|.
\end{split}
\end{align*}
We next bound each term. It follows from Lemma~\ref{Lemma_Sigma_E_tilde}(iii) and Lemma~\ref{Lemma_Sigma_E_hat}(iii) that$\|\mathbf{L}_1\| = O_{\p}(K\omega(p,T) + K/p)$.  By Lemma~\ref{Lemma_Sigma_E_tilde}(iii) and Lemma~\ref{Lemma13}(iii), we obtain that
\begin{equation*}
    \|\mathbf{L}_{2}\| = \|\mathbf{L}_{3}\| \le \big\|\widehat{\bSigma}_e^{-1} - \bSigma_e^{-1}\big\| \big\|\widehat{\bA}\big( \widehat{\bSigma}_z^{-1} + \widehat{\bA}^{\T}\widehat{\bSigma}_e^{-1}\widehat{\bA} \big)^{-1} \widehat{\bA}^{\T}\widehat{\bSigma}_e^{-1}\big\|
    = O_{\p}(K\omega(p,T) + K/p).
\end{equation*}
Applying Lemma~\ref{Lemma13}(i)--(ii), we have
\begin{flalign*}
    \|\mathbf{L}_{4}\| 
    \le& \|\widehat{\bA}\|^2 \|\bSigma_e^{-1}\|^2 \big\|\big(\widehat{\bSigma}_z^{-1}+\widehat{\bA}^{\T}\widehat{\bSigma}_e^{-1}\widehat{\bA}\big)^{-1} - \big((\bP^{\T}\bSigma_z\bP)^{-1}+\widehat{\bA}^{\T}\bSigma_e^{-1}\widehat{\bA}\big)^{-1} \big\| \\
    \le& \|\widehat{\bA}\|^2 \big\|\big(\widehat{\bSigma}_z^{-1}+\widehat{\bA}^{\T}\widehat{\bSigma}_e^{-1}\widehat{\bA}\big) - \big((\bP^{\T}\bSigma_z\bP)^{-1}+\widehat{\bA}^{\T}\bSigma_e^{-1}\widehat{\bA}\big) \big\|\\  
    &\times \big\|\big(\widehat{\bSigma}_z^{-1}+\widehat{\bA}^{\T}\widehat{\bSigma}_e^{-1}\widehat{\bA}\big)^{-1}\big\| \big\|\big((\bP^{\T}\bSigma_z\bP)^{-1}+\widehat{\bA}^{\T}\bSigma_e^{-1}\widehat{\bA}\big)^{-1} \big\| \\
    %=& O_{\p}(p/K) O_{\p}\big(p\omega(p,T) + 1\big)O_{\p}(K/p)O_{\p}(K/p)
    =& O_{\p}\big(K\omega(p,T) + K/p \big).
\end{flalign*}
Combing the above results yields that  $\big\| \widehat{\bSigma}_u^{-1} - \bSigma_u^{-1}  \big\| = O_{\p}(K\omega(p,T) + K/p),$ which completes the proof of Theorem~\ref{thm2}.
\qed

\subsection{Proof of Theorem \ref{thm3}}

\begin{lemma}\label{Lemma16}
Let $\bSigma_c = \bB\bSigma_f \bB^{\T} + \bA\bSigma_z \bA^{\T}$ and $\widehat{\bSigma}_c = \widehat{\bB}\widehat{\bSigma}_f \widehat{\bB}^{\T} + \widehat{\bA}\widehat{\bSigma}_z \widehat{\bA}^{\T}$. If Conditions~\ref{cond1}--\ref{cond6} hold and $K\omega(p,T) = o(1)$, then it holds that:
\begin{enumerate}
\item[$\mathrm{(i)}$] $\|\widehat{\bSigma}_c - \bSigma_c\| = O_{\p}(p\omega(p,T) + 1),$
\item[$\mathrm{(ii)}$] $\|\widehat{\bSigma}_c\widehat{\bSigma}_e^{-1} - \bSigma_c\bSigma_e^{-1} \| = O_{\p}(pK\omega(p,T) + K),$
\item[$\mathrm{(iii)}$] $ \lambda_{\min}(\bI_p + \widehat{\bSigma}_c\widehat{\bSigma}_e^{-1}) \ge C p$ for some constant $C > 0.$
\end{enumerate}
\end{lemma}

We are now ready to prove Theorem \ref{thm3}.

(i) Recall that $\widehat{\bSigma} = \widehat{\mathbf{B}} \widehat{\bSigma}_f \widehat{\mathbf{B}}^{\T} +  \widehat{\bSigma}_u $. Thus $\|\widehat{\bSigma} - \bSigma \|_{\bSigma}\leq\| \widehat{\mathbf{B}} \widehat{\bSigma}_f \widehat{\mathbf{B}}^{\T} - \mathbf{B} \bSigma_{f} \mathbf{B}^{\T} \|_{\bSigma} +  \|\widehat{\bSigma}_{u} - \bSigma_u \|_{\bSigma}$.  Let $\bD_T= \widehat{\bSigma}_f - \bSigma_f$. Then, using the inequality $(a+b)^2\leq 2(a^2+b^2)$ and some specific calculations, we have 
\begin{equation*}
     \| \widehat{\mathbf{B}} \widehat{\bSigma}_f \widehat{\mathbf{B}}^{\T} - \mathbf{B} \bSigma_{f} \mathbf{B}^{\T} \|^{2}_{\bSigma} \le 12\big( \|\bC_T\widehat{\bSigma}_f\bC_T\|^{2}_{\bSigma}  + \|\mathbf{B}\mathbf{D}_T\mathbf{B}^{\T} \|^{2}_{\bSigma} + \|\bC_T\widehat{\bSigma}_f\mathbf{B}^{\T} \|^{2}_{\bSigma}\big).
\end{equation*}
Applying Lemma~\ref{Lemma_CT}(i) yields that 
\begin{equation*}\|\bC_T\widehat{\bSigma}_f\bC_T\|_{\bSigma}
   = p^{-1/2}\| \bC_T \widehat{\bSigma}_f\bC_T^{\T}\bSigma^{-1} \|_{\text{F}} \le p^{-1/2}\lambda_{\max}(\bSigma^{-1})\lambda_{\max}(\widehat{\bSigma}_f) \|\bC_T \|^2_{\operatorname{F}}
   = O_{\p}(p^{1/2}\,\omega(p,T)^2).
\end{equation*}
Similar to the proof of Theorem 3.2 in \cite{fan2011high}, it is not hard to show that $\|\bB^{\T}\bSigma^{-1}\bB\| = O_{\p}(1)$. Then it follows that 
$\|\mathbf{B}\mathbf{D}_T\mathbf{B}^{\T} \|_{\bSigma}  = p^{-1/2}\|\mathbf{D}_T\mathbf{B}^{\T}\bSigma^{-1}\mathbf{B} \|_{\text{F}}
= O_{\p}(p^{-1/2} \|\mathbf{D}_T \|_{\operatorname{F}})
= O_{\p}(p^{-1/2}T^{-1/2}).$

Let $\mathbf{X} = (\mathbf{f}_1, \dots, \mathbf{f}_T)$, then $\widehat{\bSigma}_f = T^{-1}\bX\bX^{\T} - T^{-2}\bX\mathbf{1}_T\mathbf{1}_T^{\T}\bX^{\T}$, with $\mathbf{1}_T$ being a $T$-dimensional vector of ones. Note that $\|\bC_T\widehat{\bSigma}_f\mathbf{B}^{\T} \|_{\bSigma}^2 \le 8T^{-2}\|\mathbf{B}\mathbf{X}\mathbf{X}^{\T}\bC_T^{\T} \|^2_{\bSigma} +  8T^{-4}\|\mathbf{B}\mathbf{X}\mathbf{1}_T\mathbf{1}_T^{\T}\mathbf{X}^{\T}\bC_T^{\T} \|^2_{\bSigma}$. It follows from Lemma~\ref{Lemma_CT}(i) that
\begin{align*}
\|\mathbf{B}\mathbf{X}\mathbf{X}^{\T}\bC_T^{\T} \|^2_{\bSigma} 
&\le  p^{-1} \|\mathbf{X}\mathbf{X}^{\T}\bC_T^{\T}\bSigma^{-1} \|_{\text{F}} \|\bC_T \mathbf{X}\mathbf{X}^{\T}\mathbf{B}^{\T}\bSigma^{-1}\mathbf{B} \|_{\text{F}} = O_{\p}(T^{2}\omega(p,T)^2),\\
\|\mathbf{B}\mathbf{X}\mathbf{1}_T\mathbf{1}_T^{\T}\mathbf{X}^{\T}\bC_T^{\T} \|^2_{\bSigma}
&\le p^{-1} \|\mathbf{X}\mathbf{1}_T\mathbf{1}_T^{\T}\mathbf{X}^{\T}\bC_T^{\T}\bSigma^{-1} \|_{\text{F}} \|\bC_T \mathbf{X}\mathbf{1}_T\mathbf{1}_T^{\T}\mathbf{X}^{\T}\mathbf{B}^{\T}\bSigma^{-1}\mathbf{B} \|_{\text{F}}
%= O_{\p}(p^{-1})\|\bC_T \|_{\text{F}}^2 
%\|\mathbf{X}\mathbf{1}\mathbf{1}^{\T}\mathbf{X}^{\T} \|^2 
= O_{\p}(T^{4} \omega(p,T)^2).
\end{align*}
Thus, $\|\bC_T\widehat{\bSigma}_f\mathbf{B}^{\T} \|_{\bSigma} = O_{\p}(\omega(p,T))$, and then $ \| \widehat{\mathbf{B}} \widehat{\bSigma}_f \widehat{\mathbf{B}}^{\T} - \mathbf{B} \bSigma_{f} \mathbf{B}^{\T} \|_{\bSigma} = O_{\p}\big( \sqrt{p}\omega(p,T)^2 + \omega(p,T) \big).$ By Theorem~\ref{thm2}(i) and the fact that $\bSigma_u^{-1} - \bSigma^{-1}$ is a positive semi-definite matrix, it follows that
\begin{align*}
    \|\widehat{\bSigma}_{u} - \bSigma_u \|_{\bSigma} 
    &= p^{-1/2} \|\bSigma^{-1/2} (\widehat{\bSigma}_{u} - \bSigma_u ) \bSigma^{-1/2} \|_{\operatorname{F}}
    \le p^{-1/2} \|\bSigma_u^{-1/2} (\widehat{\bSigma}_{u} - \bSigma_u ) \bSigma_u^{-1/2} \|_{\operatorname{F}}=  \|\widehat{\bSigma}_{u} - \bSigma_u \|_{\bSigma_u}\\
    &    = O_{\p}(K\omega(p,T) + K/p).
\end{align*}
Thus, $ \|\widehat{\bSigma} - \bSigma \|_{\bSigma} = O_{\p}( \sqrt{p}\omega(p,T)^2 + K\omega(p,T) + K/p)$ and then part (i) holds.

(ii) By some simple calculations, we obtain that
\begin{align*}
   \| \widehat{\mathbf{B}} \widehat{\bSigma}_f \widehat{\mathbf{B}}^{\T} - \mathbf{B} \bSigma_{f} \mathbf{B}^{\T} \|_{\max} 
   \le& 2\|\bC_T \bSigma_{f}\mathbf{B}^{\T} \|_{\max}
   + \| \mathbf{B}\mathbf{D}_T\mathbf{B}^{\T}\|_{\max}
   + \|\bC_T\bSigma_{f}\bC_T^{\T}\|_{\max}\\  
   &+ 2\|\mathbf{B}\mathbf{D}_T\bC_T^{\T} \|_{\max}  + \|\bC_T\mathbf{D}_T\bC_T^{\T}\|_{\max} . 
\end{align*}
By Condition~\ref{cond6}(ii), $\|\mathbf{B} \|_{\max} = O_{\p}(1)$. Let $\mathbf{v}_i$ denote a $p$-vector with the $i$-th element being $1$ and others $0$.  Applying Lemma~\ref{Lemma_maxnorm}(iv) and Lemma~\ref{Lemma_CT}(ii) yields that
\begin{align*} 
\|\bC_T\bSigma_{f}\mathbf{B}^{\T}\|_{\max} &\le \max_{i,j\in [p]}\|\bv_i^{\T}\bC_T\bSigma_{f}\mathbf{B}^{\T}\bv_j\| 
\le \max_{i\in [p]} \|\widehat{\mathbf{b}}_i - \mathbf{b}_i\| \|\bSigma_{f}\| \max_{j\in [p]} \|\mathbf{b}_j \|
= O_{\p}\left( \omega(p,T)\right),\\
\|\mathbf{B}\mathbf{D}_T\mathbf{B}\|_{\max} &\le r^2\|\mathbf{B}\|_{\max}^2 \|\mathbf{D}_T \|_{\max} = O_{\p}(T^{-1/2}),\\
\|\bC_T\bSigma_{f}\bC_T^{\T}\|_{\max} &\le \max_{i,j\in [p]} \|\bv_i^{\T} \bC_T\bSigma_{f}\bC_T^{\T} \bv_j\| \le \max_{i\in [p]} \|\bv_i^{\T}\bC_T\|^2 \| \bSigma_{f}\| = O_{\p}(\omega(p,T)^2),\\
\|\mathbf{B}\mathbf{D}_T\bC_T^{\T}\|_{\max} &\le r^2\|\mathbf{B}\|_{\max}\|\mathbf{D}_T\|_{\max}\|\bC_T\|_{\max} = O_{\p}(T^{-1/2}\omega(p,T)),\\
\|\bC_T\mathbf{D}_T\bC_T^{\T}\|_{\max} &\le r^2 \|\mathbf{D}_T\|_{\max} \|\bC_T\|_{\max}^2 = O_{\p}(T^{-1/2}\omega(p,T)^2).
\end{align*}
Then $ \| \widehat{\mathbf{B}} \widehat{\bSigma}_f \widehat{\mathbf{B}}^{\T} - \mathbf{B} \bSigma_{f} \mathbf{B}^{\T} \|_{\max} = O_{\p}\big(\omega(p,T)\big).$ By Theorem~\ref{thm2}(ii), it follows that
\begin{equation*}
     \|\widehat{\bSigma} - \bSigma\|_{\max} \le \| \widehat{\mathbf{B}} \widehat{\bSigma}_f \widehat{\mathbf{B}}^{\T} - \mathbf{B} \bSigma_{f} \mathbf{B}^{\T} \|_{\max}  +  \|\widehat{\bSigma}_u - \bSigma_u\|_{\max}\nonumber = O_{\p}(\omega(p,T) + K/p),
\end{equation*}
which completes the proof of part (ii).

(iii) Since $\bSigma = \bSigma_c + \bSigma_e$ and $\widehat{\bSigma} = \widehat{\bSigma}_c + \widehat{\bSigma}_e$, by the triangle inequality, it follows that
\begin{flalign*}
\| \widehat{\bSigma}^{-1} - \bSigma^{-1} \|
=& \|(\widehat{\bSigma}_c + \widehat{\bSigma}_e)^{-1} + (\bSigma_c + \bSigma_e)^{-1} \|\\
\le& \|\widehat{\bSigma}_e^{-1} - \bSigma_e^{-1} \| + \|(\widehat{\bSigma}_e^{-1} - \bSigma_e^{-1})(\bI_{p} + \widehat{\bSigma}_c\widehat{\bSigma}_e^{-1})\widehat{\bSigma}_c\widehat{\bSigma}_e^{-1} \| \\
&+ \|\bSigma_e^{-1}(\bI_{p} + \widehat{\bSigma}_c\widehat{\bSigma}_e^{-1})^{-1}(\widehat{\bSigma}_c - \bSigma_c)\widehat{\bSigma}_e^{-1}\|
+ \|\bSigma_e^{-1}(\bI_{p} + \widehat{\bSigma}_c\widehat{\bSigma}_e^{-1})^{-1}\bSigma_c(\widehat{\bSigma}_e^{-1} - \bSigma_e^{-1})\| \\
&+ \|\bSigma_e^{-1}\{ (\bI_{p} + \widehat{\bSigma}_c\widehat{\bSigma}_e^{-1})^{-1} - (\bI_{p} + \bSigma_c \bSigma_e^{-1})^{-1} \}    \bSigma_c \bSigma_e^{-1}  \| \\
:=& \|\mathbf{L}_{5}\| + \|\mathbf{L}_{6}\| + \|\mathbf{L}_{7}\| + \|\mathbf{L}_{8}\| + \|\mathbf{L}_{9}\|.
\end{flalign*}
Clearly, $\|\mathbf{L}_{5}\| = O_{\p}(K\omega(p,T) + K/p)$ by Lemma~\ref{Lemma_Sigma_E_tilde}(iii) and Lemma~\ref{Lemma_Sigma_E_hat}(iii). Applying Lemma~\ref{Lemma16}(iii), we obtain that  $\|\mathbf{L}_{6}\| \le \|\widehat{\bSigma}_e^{-1} - \bSigma_e^{-1} \| \|\bI_{p} + \widehat{\bSigma}_c\widehat{\bSigma}_e^{-1}\| \|\widehat{\bSigma}_c\| \|\widehat{\bSigma}_e^{-1} \| =  O_{\p}(K\omega(p,T) + K/p).$ Moreover, by Lemma~\ref{Lemma16}(i) and (iii), it follows that $\|\mathbf{L}_{7}\| \le \|\bSigma_e^{-1}\| \|(\bI_{p} + \widehat{\bSigma}_c\widehat{\bSigma}_e^{-1})^{-1}\| \|\widehat{\bSigma}_c - \bSigma_c\| \|\widehat{\bSigma}_e^{-1}\| = O_{\p}(\omega(p,T) + 1/p).$ Applying Lemma~\ref{Lemma_Sigma_E_tilde}(iii), Lemma~\ref{Lemma_Sigma_E_hat}(iii) and Lemma~\ref{Lemma16}(iii) yields that $\|\mathbf{L}_8\| \le \|\bSigma_e^{-1}\| \|(\bI_{p} + \widehat{\bSigma}_c \widehat{\bSigma}_e^{-1})^{-1}\| \|\bSigma_c\| \|(\widehat{\bSigma}_e^{-1} - \bSigma_e^{-1})\| = O_{\p}(K\omega(p,T) + K/p).$ For $\|\mathbf{L}_{9}\|$, it follows from Lemma~\ref{Lemma16}(ii)--(iii) that
\begin{align*}
\|\mathbf{L}_{9}\| 
&\le \|\bSigma_e^{-1}\| \|(\bI_{p} + \widehat{\bSigma}_c\widehat{\bSigma}_e^{-1})^{-1} - (\bI_{p} + \bSigma_c \bSigma_e^{-1})^{-1}\|    \|\bSigma_c\| \|\bSigma_e^{-1}\| \\
&\le \|(\bI_{p} + \widehat{\bSigma}_c\widehat{\bSigma}_e^{-1})^{-1}\|
\|\widehat{\bSigma}_c\widehat{\bSigma}_e^{-1} -  \bSigma_c \bSigma_e^{-1}\| 
\|(\bI_{p} + \bSigma_c \bSigma_e^{-1})^{-1} \| 
\|\bSigma_c\| \|\bSigma_e^{-1}\|^2   
    = O_{\p}(\omega(p,T) + 1/p).
\end{align*}
Thus, $\|\widehat{\bSigma}^{-1} - \bSigma^{-1} \| = O_{\p}\big(K \omega(p,T) + K/p\big),$ which completes the proof of part (iii).

\end{appendix}

\linespread{1.5}\selectfont
\bibliographystyle{dcu}
\bibliography{ref}

\end{document}